\documentclass[structabstract]{aa} 
\usepackage{graphicx} 
\usepackage{txfonts} 
\usepackage{natbib} 
\usepackage{longtable} 
\usepackage{lscape} 

\bibpunct{(}{)}{;}{a}{}{,}

\newcommand{\mum}{\,$\mu$m} 
\newcommand{\degree}{$^{\circ}$} 
\newcommand{\Msun}{$M_{\odot}$} 
 
\newcommand{\Msunyr}{$M_{\odot}$yr$^{-1}$}

\newcommand{\sek}{Sect.} 
\newcommand{\fig}{Fig.} 
\newcommand{\tab}{table} 
\newcommand{\fid}{$f_{\rm{id}}$} 
\newcommand{\rev}{ } 
\newcommand{\newrev}{ } 
 \newcommand{\newnewrev}{ } 
\begin{document} 
\title{Star formation and disk properties in Pismis\,24\thanks{Based on observations performed at ESO's La Silla-Paranal observatory under programme 081.C-0467.}} 
 
\author{M.~Fang\inst{1,2} \and  R.~van~Boekel\inst{1} \and R.\,R.~King\inst{3} \and Th.~Henning\inst{1} \and J.~Bouwman\inst{1} \and Y.~Doi\inst{4} \and Y.~K.~Okamoto\inst{5} \and V.~Roccatagliata\inst{1,6} \and A.~Sicilia-Aguilar\inst{1,7}}  
\institute{Max-Planck Institute for Astronomy, K\"onigstuhl 17, 
           D-69117 Heidelberg, Germany 
           \and 
           Purple Mountain Observatory, 2 West Beijing Road, 210008 Nanjing, China
           \and 
           School of Physics, University of Exeter, Stocker Road, Exeter EX4 4QL
           \and 
           Graduate School of Arts and Sciences, University of Tokyo, Japan. 
           \and 
           2-1-1 Bunkyo, Mito, Ibaraki, Japan 
           \and 
          Space Telescope Science Institute, 3700 San Martin Drive, Baltimore MD 21210, USA 
          \and
        Departamento de F\'{\i}sica Te\'{o}rica, Universidad Aut\'{o}noma de Madrid, Cantoblanco 28049, Madrid, Spain
         }     
 
   \date{Received 12 October 2010; accepted 27 December 2011}

\abstract{
Circumstellar disks are expected to evolve quickly in massive young clusters harboring many OB-type stars. Two processes have been proposed to drive the disk evolution in such cruel environments: (1) gravitational interaction between circumstellar disks and nearby passing stars (stellar encounters), and (2) photoevaporation by UV photons from massive stars. The relative importance of both mechanisms is not well understood. Studies of massive young star clusters can provide observational constraints on the processes of driving disk evolution.} 
{We investigate the properties of young stars and their disks in the NGC\,6357 complex, concentrating on the most massive star cluster within the complex: Pismis\,24.} 
{We use infrared data from the 2MASS and Spitzer GLIMPSE surveys, complemented with our own deep Spitzer imaging of the central regions of Pismis\,24, in combination with X-ray data to search for young stellar objects (YSOs) in NGC\,6357 complex. The infrared data constrain the disk presence and are complemented by optical photometric and spectroscopic observations, obtained with VLT/VIMOS, that constrain the properties of the central stars. For those stars with reliable spectral types we combine spectra and photometry to estimate the mass and age. For cluster members without reliable spectral types we obtain the mass and age probability distributions from $R$ and $I$-band photometry, assuming these stars have the same extinction distribution as those in the ``spectroscopic'' sample. We compare the disk properties in the Pismis\,24 cluster with those in other clusters/star-forming regions employing infrared color-color diagrams.} 
{We discover two new young clusters in the NGC\,6357 complex. We give a revised distance estimate for Pismis\,24 of 1.7$\pm$0.2\,kpc. We find that the massive star Pis\,24-18 is a binary system, with the secondary being the main X-ray source of the pair. We provide photometry in 9 bands between 0.55 and 9~\mum \ for the members of the Pismis\,24 cluster. We derive the cluster mass function and find that up to the completeness limit at low masses it agrees well with the initial mass function of the Trapezium cluster. We derive a median age of 1\,Myr for the Pismis\,24 cluster members. We find five proplyds in HST archival imaging of the cluster, four of which are newly found. In all cases the proplyd tails are pointing directly away from the massive star system Pis\,24-1. One proplyd shows a second tail, pointing away from Pis\,24-2, suggesting this object is being photoevaporated from two directions simultaneously. We find that the global disk frequency ($\sim$30\%) in Pismis\,24 is much lower than some other clusters of similar age, such as the Orion Nebula Cluster. When comparing the disk frequencies in 19 clusters/star-forming regions of various ages and different (massive) star content, we find that the disks in clusters harboring extremely massive stars (typically earlier than O5), like Pismis\,24, are dissipated roughly twice as quickly as in clusters/star-forming regions without extremely massive stars. Within Pismis\,24, we find that the disk frequency within a projected distance of 0.6\,pc from Pis\,24-1 is substantially lower than at larger radii ($\sim$19\% vs. $\sim$37\%). We argue for a combination of photoevaporation and irradiation with ionizing UV photons from nearby massive stars, causing increased MRI-induced turbulence and associated accretion activity, to play an important role in the dissipation of low-mass star disks in Pismis~24.} 
{}

 
 \maketitle

\section{Introduction} 
The influence of nearby massive young stars on the evolution of circumstellar disks is still not well understood. Spitzer observations have revealed a clear anti-correlation between the frequencies of circumstellar disks and the presence of massive stars in these clusters \citep[e.g., NGC\,2244, NGC\,6611][]{2007ApJ...660.1532B,2009A&A...496..453G}. UV irradiation by hot, massive stars causing photoevaporation of disks around neighboring lower mass young stars is a favored mechanism to explain the observed low disk frequencies near massive stars. In addition, massive stars preferentially reside in the centers of clusters \citep{2007ARA&A..45..481Z} where stellar densities are extremely high. In such environments stellar encounters, causing gravitational interaction between the circumstellar disks and nearby cluster members, are proposed to play a role in disk dissipation \citep{2006A&A...454..811P,2010A&A...509A..63O}. The relative importance of both mechanisms for the dissipation of circumstellar disks is not well constrained. Young clusters harboring very massive stars need to be studied to shed light on this issue, Pismis\,24 constitutes a well suited example of such clusters. 
 
Pismis\,24 is located in the Sagittarius spiral arm and contains dozens of massive OB-type stars, with two extremely luminous members: Pis\,24-1 (O3\,I) and Pis\,24-17 (O3.5\,III) \citep{2001AJ....121.1050M}. High-resolution observations have resolved Pis\,24-1 into a compact hierarchical triple system consisting of Pis\,24-1 NE (unresolved spectroscopic binary) and Pis\,24-1 SW \citep{2007ApJ...660.1480M}. Pis\,24-1\,NE and Pis\,24-1\,SW have an estimated mass of around 100\,\Msun \ each. The distance to Pismis\,24 has been estimated in different ways. \citet{1970A&A.....6..364W} derive a kinematic distance to Pismis\,24 of 1.0$\pm$2.3\,kpc. \citet{1978A&A....69...51N} obtain a distance of 1.74$\pm$0.31\,kpc to Pismis\,24 using a color-magnitude diagram. Employing the spectroscopic parallax method, \citet{2001AJ....121.1050M} derive a distance of 2.56$\pm$0.10\,kpc for Pismis\,24. In the current work we re-address the distance to Pismis\,24 based on the positions of the most massive members in the Hertzsprung-Russell diagram (see \sek~\ref{sec:pis24_distance}) and find that the most likely distance is 1.7\,kpc. We adopt this value throughout the current paper. 
 
Our inventory of the stellar content of Pismis\,24 has long been limited to the massive members \citep{1970A&A.....6..364W,2001AJ....121.1050M}. Observations with the Chandra X-ray telescope have dramatically improved this situation and allowed hundreds of lower mass members to be identified \citep{2007ApJS..168..100W}. A total of 779 X-ray sources have been found in the Pismis\,24 region, of which 616 sources have associated optical or infrared counterparts. With an estimated age of $\sim$1\,Myr \citep{2001AJ....121.1050M} Pismis\,24 is an extremely interesting laboratory for investigating the circumstellar disk evolution, with the similarly old Orion Nebula Cluster as a local reference. 
 
The Pismis\,24 cluster resides within NGC\,6357, a large complex of extended nebulosity. Within NGC\,6357 there are three known H{\scriptsize II} regions at different evolutionary stages: G353.2+0.9, G353.2+0.7, and G353.1+0.6 \citep{1990A&A...232..477F}. G353.2+0.9, located near the Pismis\,24 cluster, is the youngest and brightest region and was further resolved into three compact H{\scriptsize II} regions using the high-resolution VLA observations \citep{1990A&A...232..477F}. Between G353.2+0.9 and the Pismis\,24 cluster there is an ionization front shielding the cloud material from most of the UV photons emitted from massive stars in the cluster. It is therefore possible that G353.2+0.9 is ionized by its own, internal sources \citep{1990A&A...232..477F}. Indeed, infrared observations have revealed several embedded objects in G353.2+0.9 \citep{1986A&A...170...97P,1990A&A...232..477F,2007ApJS..168..100W}. G353.1+0.6 is a more evolved H{\scriptsize II} region which is expanding and interacting with molecular cloud material on its northern side \citep{1990A&A...232..477F,1997A&A...320..972M}. G353.2+0.7 is the most evolved and diffuse H{\scriptsize II} region and shows no compact components \citep{1979AuJPA..48....1H,1990A&A...232..477F}.  
 
A dust continuum emission survey at 1.2\,mm has revealed {\newrev 73} dense cores in the NGC\,6357 region, of which {\newrev 5} have masses above 200\,\Msun \ \citep{2007ApJ...668..906M,2010A&A...515A..55R}. Follow-up observations searching for molecular line emission in {\newrev 12} high-mass dense cores ($\ge$100\,\Msun) show that all of them belong to the NGC\,6357 complex. Among these, {\newrev 6} dense cores were identified as starless cores \citep{2010A&A...515A..55R}. 
 
In this paper we will first investigate the star formation activity in the NGC\,6357 complex as a whole, and then focus on the stellar and disk properties of the members of the central cluster Pismis\,24. We arrange this paper as follows. In section~\ref{Sec:data} we describe the observations and data reduction. In section~\ref{Sec:result} we present our results which are then discussed in section~\ref{Sec:discussion}. We summarize our effort in section~\ref{Sec:summary}. 
 
\section{Observations and data reduction}\label{Sec:data} 
\begin{figure} 
\centering 
\includegraphics[width=\columnwidth]{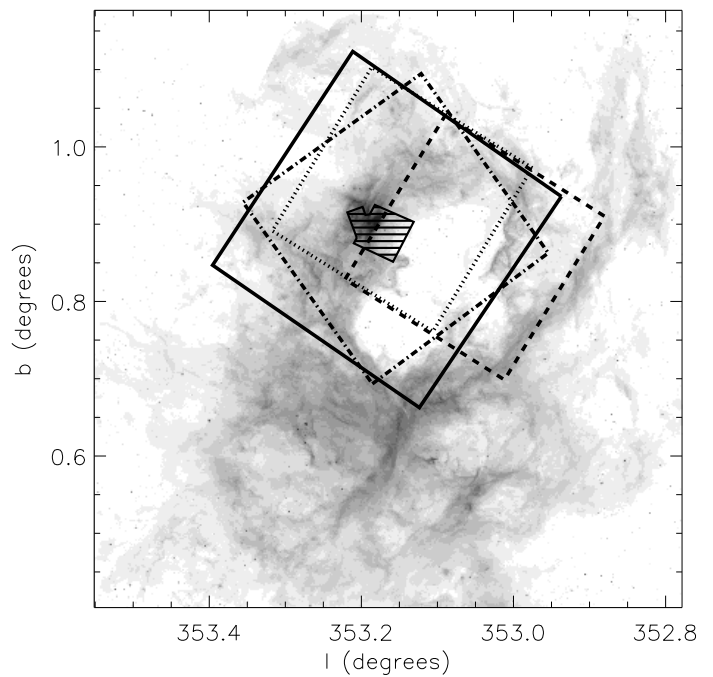} 
\caption{{\newrev The FOVs of different observations used in this paper. The background: 8.0\,\mum (GLIMPSE). The FOVs of our deep IRAC imaging are shown with the dashed lines ([3.6] and [5.8] bands) and dotted lines ([4.5] and [8.0] bands). The dash-dotted lines present the FOV of the Chandra X-ray observation. The solid lines show the FOVs for VIMOS imaging and spectroscopy. The central line-filled regions were covered with HST observations.}}\label{Fig:FOV} 
\end{figure}

This study is based on a large collection of observational data. We use photometric data at optical, infrared and X-ray wavelengths, as well as spectroscopy in the 4800 to 10000\,$\AA$ range. The field of view (FOV) for each of the individual datasets is shown in Fig~\ref{Fig:FOV}. 
 
\begin{figure*} 
\centering 
\includegraphics[width=13cm]{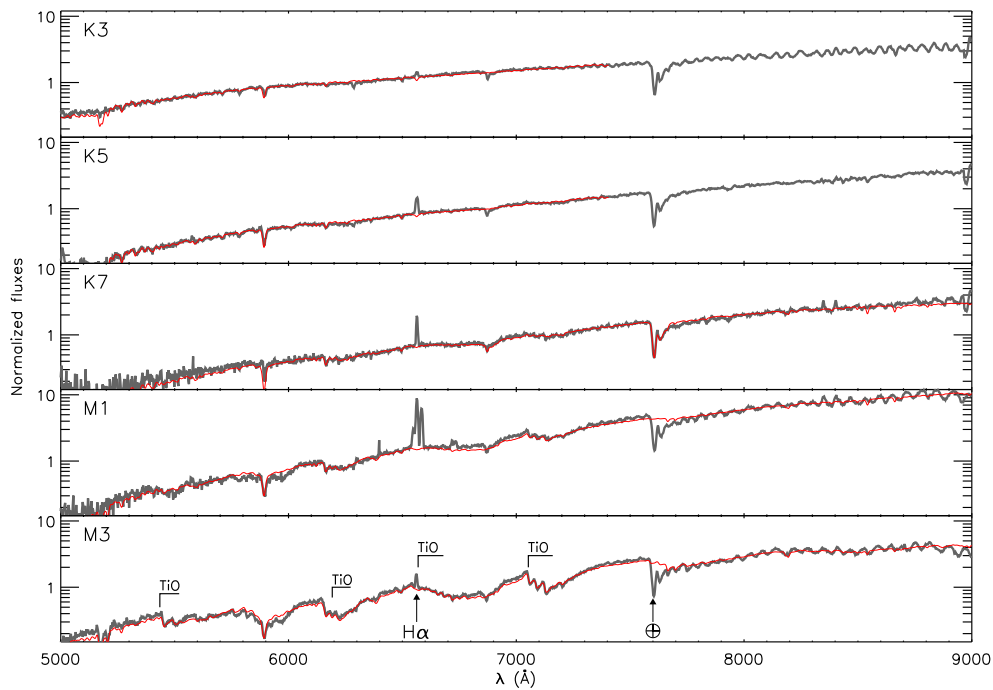} 
\caption{Typical spectra (grey thick lines) from our VIMOS observations covering the range of spectral types from K3 to M3. Spectral templates (thin lines) with the same spectral types and spectral resolution are also shown. The templates have been reddened to fit our VIMOS spectra.  The prominent H$\alpha$ emission line and TiO  absorption features are indicated. The wiggles on the spectra beyond 7600$\AA$ are instrumental artifacts due to the fringing.}\label{Fig:example_st} 
\end{figure*}

\subsection{Infrared photometry} 
  Near-infrared photometry in the $J$, $H$, and $K_{\rm s}$ bands was taken from the Two-Micron All Sky Survey \citep[2MASS][]{2006AJ....131.1163S}. Mid-infrared photometry at 3.6, 4.5, 5.8, and 8.0\mum\ obtained with the Spitzer Space Telescope IRAC camera \citep{2004ApJS..154...10F} was taken from the the Galactic Legacy Infrared Mid-Plane Survey Extraordinaire \citep[GLIMPSE\,I][]{2003PASP..115..953B} survey, and supplemented with our own deep observations (Program ID 30726) in the central cluster regions. 
 
{\newrev \subsubsection{2MASS survey} 
The 2MASS survey imaged  the entire sky in the $J$, $H$, and $K_{\rm s}$ bands. We extracted the photometry of all the point sources in NGC\,6357 complex. The  typical 3$\sigma$ limits for the 2MASS survey are 17.1, 16.4 and 15.3\,mag in $J$, $H$, and $K_{\rm s}$, respectively \citep{2003tmc..book.....C}.}

\subsubsection{GLIMPSE survey} 
The GLIMPSE~I survey covers the galactic plane (10\degree\,$<|l|<$65\degree, $|b|<$1\degree\,) with imaging in the four IRAC bands (3.6, 4.5, 5.8, and 8.0 \mum).  We adopted the photometry of all the point sources in NGC\,6357 complex as given in the GLIMPSE {\newrev catalogue}, with photometric uncertainties below 0.2\,mag in all four IRAC bands. {\newrev The typical 3$\sigma$ limits for GLIMPSE survey are 15.5, 15.0, 13.0, and 13.0\,mag at the 3.6, 4.5, 5.8, and 8.0\,\mum, respectively \citep{2009PASP..121..213C}. In the PISMIS~24 field, where the number density of stars is high, the 3$\sigma$ limiting magnitudes in regions without strong nebular background are estimated to be $\sim$15.2, 14.8, 13.0, and 12.6\,mag in the four IRAC bands, respectively.}  
 
\subsubsection{Deep IRAC imaging of the central Pismis\,24 region} 
We have performed deep observations towards the Pismis\,24 cluster with the Spitzer IRAC camera. The observations have been done on {\rev September 29, 2006}, with exposure times of 0.4\,s and 10.4\,s. {\newrev We mosaiced images using the SSC mosaic software MOPEX}, and performed PSF photometry using the IDL codes described in \cite{2009A&A...504..461F}. We compared our photometry with the values given in the GLIMPSE catalogue for common sources and found small systematic differences between both data sets of -0.04, -0.007,-0.04,-0.1 magnitudes in the [3.6], [4.5], [5.8], and [8.0] bands, respectively.  These difference may be due to the slightly different psf-fitting models. We applied the corresponding scaling factors to our deep imaging such that it has the same absolute flux levels as the GLIMPSE data. {\newrev The 3$\sigma$ limiting magnitudes for our images  are estimated to be $\sim$15.7, 15.5, 14.0, and 13.0\,mag, respectively, in the four IRAC bands, in the region {\newnewrev in the} absence of  nebulosity.}

\subsection{Optical photometry} 
We have imaged the cluster Pismis\,24 in the $R$ and $I$-band filters using the Visible Wide Field Imager and Multi-Object Spectrograph \citep[VIMOS,][]{2003SPIE.4841.1670L} at the ESO VLT. The $R$-band observations were performed on 2008 April 1 and 6, and the $I$-band observations were done on 2008 May 1. In order to increase dynamic range we took the images with five exposures (1, 14, 45, 150, and 300 seconds) for every pointing. We performed standard data reduction for optical imaging consisting of bias subtraction and flat-fielding. We then performed PSF photometry on the reduced images, taking into account that the PSF shows substantial variations in shape over the VIMOS FOV. Instead of using a single PSF for a given observation, we divided each image into 16 sub-regions. In each sub-region we extracted a PSF model from the isolated stars, and used it to do PSF fitting for each star in the sub-region. For every star that was detected in multiple exposures of different integration time, we adopt the photometry from the longest exposure in which the peak level of the star remains in the linear regime of the CCD. We calibrated the $R$-band photometry using observations of the standard fields SA\,110 (for data taken on 2008 April 1) and Rubin 149 (for data taken on 2008 April 6). The $I$-band photometry was calibrated using standard stars observed in the PG\,0918 field \citep{2000PASP..112..925S}.

\begin{figure*} 
\centering 
\includegraphics[width=11cm]{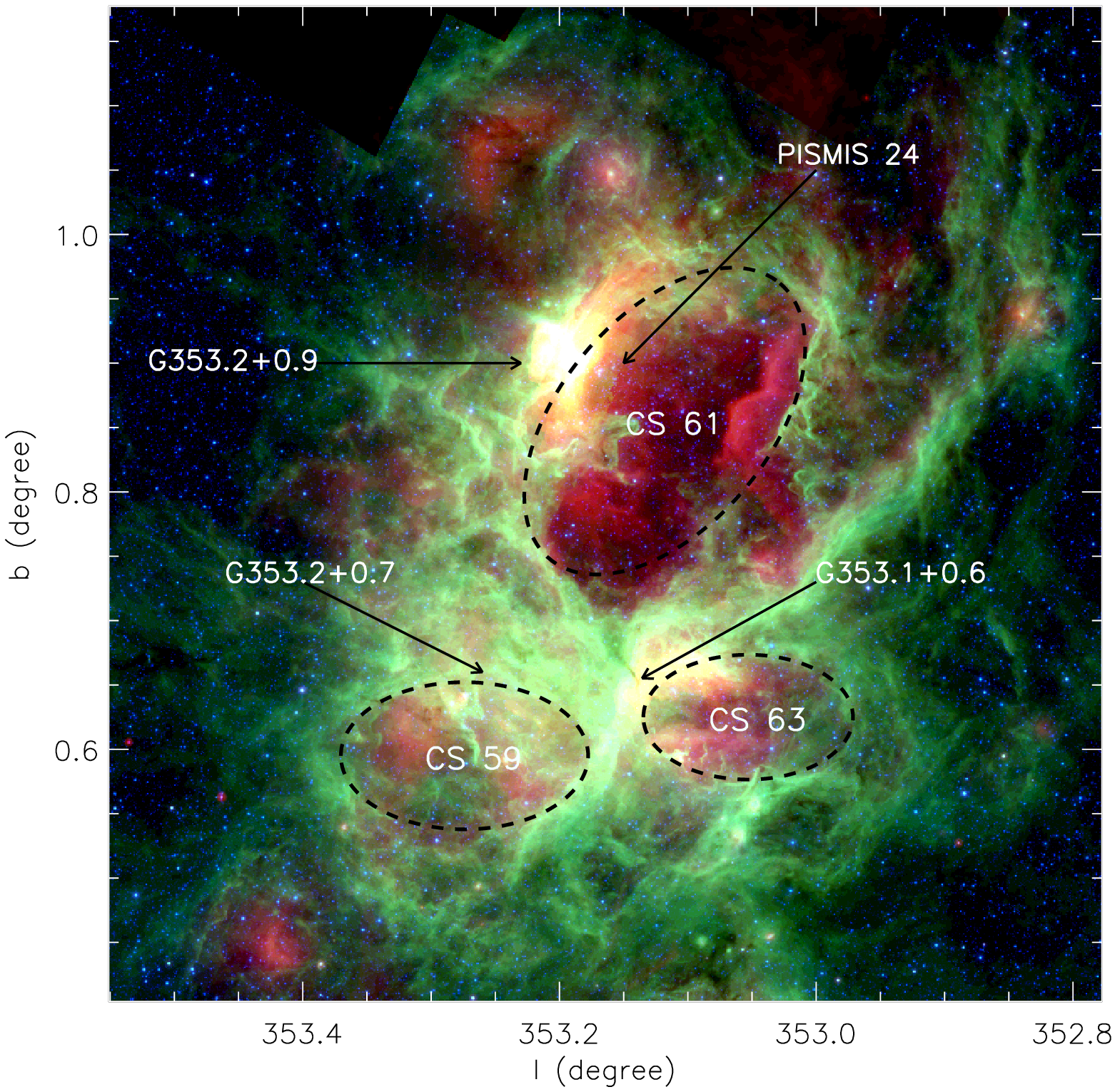} 
\caption{Spitzer color image of NGC\,6357 complex(red: 24\mum, green: 8.0\mum, and blue: 4.5\mum). The  4.5\mum\, and 8.0\mum\, images are from GLIMIPSE survey. The  24\mum\ image is a combination of the images from  MIPSGAL and MSX surveys. The three H{\scriptsize II} regions G353.2+0.9, G353.2+0.7, and G353.1+0.6 are labeled. The position of the Pismis\,24 cluster is marked. The dashed lines depict the inner rims of the three bubbles identified by \citet{2007ApJ...670..428C}.}\label{Fig:Glimpse_color_map} 
\end{figure*}

\subsection{X-ray source catalogue} 
Pismis\,24 cluster has been observed with the Imaging Array of the Advanced CCD Imaging Spectrometer (ACIS-I) mounted on the Chandra space telescope. In this region, 779 X-ray sources are detected \citep{2007ApJS..168..100W}. In this paper, we match X-ray sources to the sources detected in VIMOS $R$ and $I$ bands based on positional coincidence, using a 1\farcs5 tolerance. The photometry in 9 bands  between 0.55 and 9~\mum \ for counterparts of X-ray sources are listed in \tab~\ref{tab:photometry}.  Since Pismis\,24 is located in the direction of the galactic center the density of unrelated background sources is high, which complicates establishing cluster membership. {\newrev Young stars can be identified due to their X-ray emission, which is enhanced in comparison to field stars. The X-ray emission from high mass stars is thought to {\newnewrev arise} in shocks in their fast, radiatively driven winds. Lower mass stars produce X-ray emission due to magnetic reconnection flares similar to those seen on the solar surface, but with X-ray fluxes 2 to 3 orders of magnitude higher than seen in the field population. In this work, we will use this X-ray emission to identify young stars associated with the young stellar cluster Pismis 24. The potential contamination of our sample with X-ray bright AGN and foreground stars was shown to be $\lesssim$4\%  \citep{2007ApJS..168..100W} and is thus not a major concern.}

  Since the central regions of the Pismis 24 cluster have a high space density of sources and the accuracy of the optical and X-ray positions is limited to typically 0\farcs6 and 1\farcs0, respectively, matching the optical and X-ray positions requires care. Choosing a large matching radius ensures that all real optical and X-ray pairs are matched but may also result in substantial numbers of ``false positives'', i.e. a match between physically unrelated optical and X-ray sources. Choosing a small matching radius would cause many physically associated pairs to be lost from the analysis. We chose 1\farcs5 as a compromise. To test for the number of potential false positives, we applied a positional shift of 25$''$ to the optical positions and then matched those to the X-ray catalog. This indeed yields a fair number of false matches: roughly 1/6 of the number of matches in the original catalog. Thus, strictly speaking, statistically about 1/6 of our sources could be false positives. This, however, is a very pessimistic estimate: for individual sources it would mean that there is an optical source near the X-ray position that is unrelated to X-ray source and that the true X-ray source counterpart is much fainter in the optical than the ``false match''. This situation will not occur often, though for individual cases it cannot, of course, be excluded with certainty.

\begin{figure*} 
\centering 
\includegraphics[width=11cm]{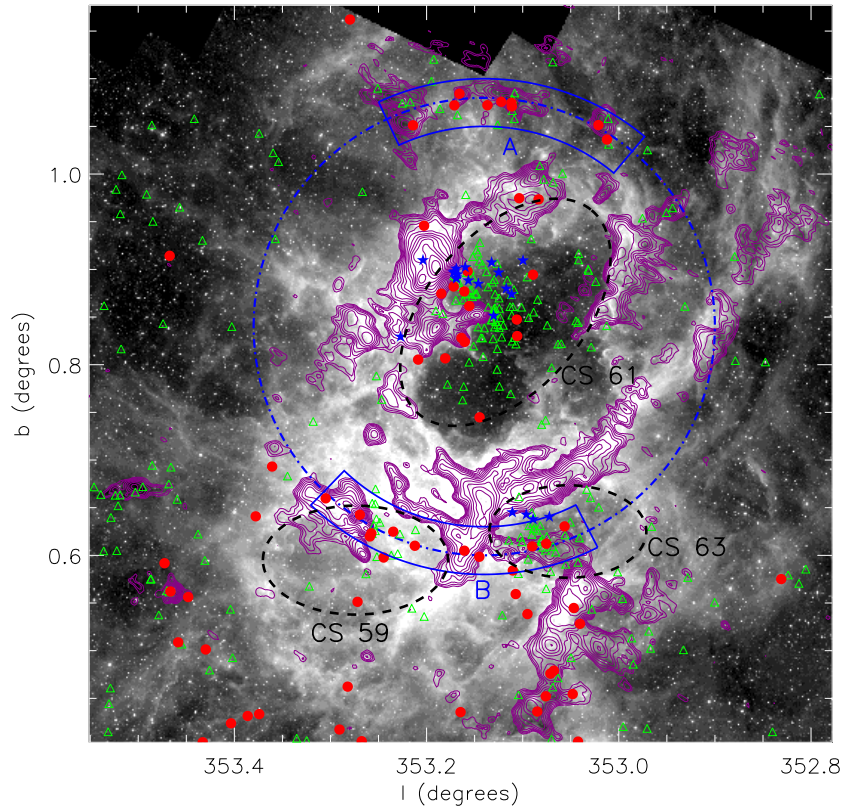} 
\caption{{\newrev 1.2}\,mm emission (contours) over-plotted on the IRAC [5.8] image of the NGC\,6357 complex. {\newrev The beam size of 1.2\,mm emission is 24$''$}. The red filled circles mark the Flat/class\,I objects, and the open triangles mark the class\,II objects. The asterisks indicate  known massive stars. The dashed lines depict the inner rims of the three bubbles identified by \citet{2007ApJ...670..428C}. In regions A and B, the YSO distribution shows arcs centered on the bubble CS\,61. }\label{Fig:mm_map} 
\end{figure*}

\subsection{Optical spectroscopy} 
Spectroscopic observations of the Pismis\,24 cluster were performed with VIMOS in August 2008.  {\newrev The targets for the VIMOS spectroscopy were selected from the sources found in the VIMOS pre-imaging with a match (within 1\farcs0) to an X-ray source from the catalogue of \citet{2007ApJS..168..100W}. Once as many of these as possible were accommodated by the slit mask, additional targets were added where space allowed to include stars with fluxes in the range 11$<$J$<$14 in an attempt to identify some intermediate mass stars which are least sensitive to X-ray detection \citep{2003ApJ...584..911F}.} We used a low resolution grism combined with order sorting filter GG475 and a slit width of 1\farcs0, resulting in an effective spectral resolution of $\lambda/\delta\lambda$=580 and a spectral coverage from 4800$\AA$ to 10000$\AA$. The spectroscopic data were reduced with the VIMOS pipeline, which was also used to extract the raw spectra of the individual targets. These were then flux-calibrated using observations of the standard star LTT\,6248. 
 
In the following we will describe how we do the spectral classification of our stars using the VIMOS spectra. We have adopted a two-step approach. In step one we estimate the spectral types from our spectra using the IDL package 'Hammer' \citep{2007AJ....134.2398C}. The Hammer code was originally designed to classify spectra of stars in the absence of appreciable extinction. It uses 29 spectral ``features'' in the classification process, among which two (``BlueColor'' and ``Color-1'') are related to the broad-band spectral continuum slope. The latter cannot be used for stars whose spectra are reddened due to absorption by intervening dust. The remaining 27 diagnostics are ``narrow-band'' features, and not affected by extinction. In order to make the Hammer code suitable for classifying our young stars, that typically suffer several magnitudes of optical extinction, we removed the BlueColor and Color-1 features from the Hammer code, keeping the remaining 27 diagnostics. We then ran the code to classify all our spectra, giving initial estimates of the spectral types. In the second step of the classification process we tested the results by fitting spectral templates with spectral types as derived using the Hammer code to the observed VIMOS spectra. {\rev The spectral templates are from the 'Hammer' package.} We have two free parameters in this fit: the $V$-band extinction by which the spectral template is reddened using a standard ISM extinction law {\rev \citep{1989ApJ...345..245C}}, and the scaling factor calibrating the spectral template to the observed absolute flux level. The results were then visually inspected for each star to ensure a satisfactory fit to the data was obtained. When the spectral template with initial spectral type from ``Hammer'' code could not properly fit the observations, the input spectral type was varied by several subclasses until we obtained a good match between the observed spectrum and the spectral template. The spectral type of the best-fitting template was finally adopted as the spectral type of each respective object. In Fig~\ref{Fig:example_st} we show examples of our VIMOS spectra with a range of spectral types from early K to early M, representing the majority of the young stars in our sample. In each case, we have overplotted the best-fitting spectral template. Over this wavelength range, the changes in the spectral shape are clearly visible, in particular the strength of the TiO absorption bands is a prominent diagnostic for late K to M type stars.

\subsection{Complementary data sets} 
A part of the Pismis\,24 star cluster has been observed with the Hubble space telescope in the F502N, F656N, F673N, and F850LP bands\footnote{We obtained these HST images from http://hla.stsci.edu/hlaview.html.}. We have used these HST images to search for proplyds, i.e. young stars whose circumstellar disks are being photoevaporated by UV photons from nearby massive stars, resulting in a head-tail shaped appearance (see \sek~\ref{Sec:disk}(a)).  
 
The 24\,\mum \ image of the NGC\,6357 complex was taken from the Spitzer MIPSGAL survey \citep{2009PASP..121...76C}. Parts of the 24\,\mum \ Spitzer data were saturated, we fill in those regions using 21.3\mum \ data from the Midcourse Space Experiment \citep[MSX,][]{2001AJ....121.2819P} survey. We tied the flux scale of the MSX image to that of the MIPSGAL observations using the common unsaturated regions, ignoring the relatively minor differences in the spectral response between both images.

Imaging at 1.2\,mm was performed with the SEST Imaging Bolometer Array \citep{2007ApJ...668..906M} and traces the dust continuum emission in high column density regions \citep[$A_{\rm v}\ge$15\,mag, ][]{2010A&A...515A..55R}. These data were used to {\newrev locate the dense molecular regions} in NGC\,6357 complex. 
 
\section{Results}\label{Sec:result}

\subsection{The NGC\,6357 complex} 
In this section, we will present a global view of NGC\,6357. We will show the dust emission from mid-infrared to millimeter wavelengths and then investigate the star formation activity throughout the region. 
 
 \subsubsection{Dust emission in NGC\,6357} 
In \fig~\ref{Fig:Glimpse_color_map} we show a three-color composite of the whole NGC\,6357 complex (4.5, 8.0, and 24~\mum \ in blue, green, and red, respectively), using data from the GLIMPSE survey at 4.5 and 8.0~\mum), and the MIPSGAL surveys at 24\,\mum \ with the saturated regions replaced with the data from MSX survey at 21.3\,\mum. The emission in the 4.5\,\mum \ IRAC band traces  {\newrev the stars that appear as point sources, some diffuse emission emitted or scattered by dust, and Br$\alpha$ emission. The 8.0\,\mum \ and  5.8\,\mum\ IRAC bands are dominated by the emission from  Polycyclic Aromatic Hydrocarbons (PAHs) \citep{2007ApJ...660..346P},  tracing the surface of molecular clouds. The 24\,\mum\ emission is dominated by continuum dust emission.}

The most striking aspect of \fig~\ref{Fig:Glimpse_color_map} is that the whole region is full of filamentary diffuse emission, with some ``bubbles'' where {\newrev PAH emission} is faint or absent and 24\,\mum \ emission is dominating. In this field \citet{2007ApJ...670..428C} identified three bubbles, named CS\,59, CS\,61, and CS\,63, of which CS\,61 is the biggest. The bright H{\scriptsize II} region G353.2+0.9 lies near bubble CS\,61 \citep{1990A&A...232..477F}. The Pismis\,24 cluster appears to be located within CS\,61 and the strong UV field and stellar winds from the massive cluster members are likely responsible for creating CS\,61. The absence of 8.0\,\mum \ emission within CS\,61 and the ``sharp'' edges of the bubble at this wavelength (\fig~\ref{Fig:Glimpse_color_map}) can be explained by the absence of PAHs within the bubble due to their destruction by extreme ultraviolet (EUV) photons from massive stars \citep{1992MNRAS.258..841V}.  Surrounding CS\,61 there are high density shells which are traced by the dust continuum emissions at 1.2\,mm (see Fig~\ref{Fig:mm_map}). Here, the EUV photon fluxes have already been diminished due to absorption by gas and dust within the bubble and PAHs can survive. There are still sufficient amounts of UV photons of lower energy to excite the PAH molecules, which then glow brightly and mark the boundary between the dense cloud material and the cavities carved by the star clusters. 
 
The bubbles CS\,59 and CS\,63 are much smaller than CS\,61. Near these bubbles the H{\scriptsize II} regions G353.2+0.7 and G353.1+0.6, respectively, are located. Inside bubble CS\,63 there are four known OB stars \citep{1984A&A...137...58N}, which are ionizing the H{\scriptsize II} region G353.1+0.6 \citep{1990A&A...232..477F} and have likely created this bubble. Inside the bubble CS\,59 there are no known massive stars in the literature. Further observations are required to understand the origin of this bubble. In contrast to bubble CS\,61, there is weak 8.0\mum \ emission inside bubbles CS\,59 and CS\,63, suggesting that their central stars  cannot emit sufficient EUV photons to destroy all the PAHs inside the bubbles. 
 
In \fig~\ref{Fig:mm_map} we show the {\newrev 1.2}\,mm dust continuum emission with contours, overplotted on the 5.8\,\mum \ IRAC image.  {\newrev The millimeter data, tracing the molecular clouds, tend to be concentrated in ring-like structures surrounding the bubbles CS\,61, CS\,59, and CS\,63.}  Similar structures of molecular gas have been found around many mid-infrared bubbles \citep{2009A&A...496..177D,2010ApJ...709..791B}. They are thought to arise due to the compression of the molecular clouds by the expanding shock fronts produced by stellar winds or the pressure-driven expansion of the H{\scriptsize II} regions \citep{1975ApJ...200L.107C,2003ApJ...594..888F}.

\subsubsection{Star formation in NGC\,6357}\label{Sec:cluster} 
 
Using the infrared data from the 2MASS and GLIMPSE surveys we can make an inventory of the disk-bearing young star population in the whole NGC\,6357 complex, and thus investigate the global recent star formation activity. The infrared excess emission due to the dusty circumstellar disks causes the infrared colors of stars with disks to be distinctly different from those of diskless objects. {\newrev However, young cluster members that have already lost their disks cannot be robustly distinguished from unrelated field objects based on infrared colors alone. Therefore, in this section, we restrict ourselves to studying the disk-bearing stars only to trace recent star forming activity in NGC\,6357 complex.}

\begin{figure} 
\centering 
\includegraphics[width=0.85\columnwidth]{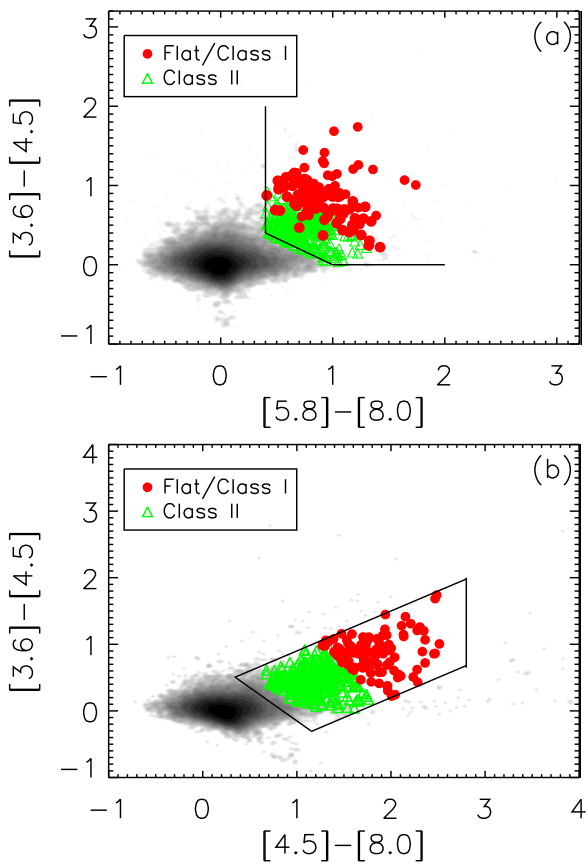} 
\caption{(a) Spitzer/IRAC [3.6]-[4.5] vs. [5.8]-[8.0]  and (b) [3.6]-[4.5] vs. [4.5]-[8.0] color-color diagrams. YSO candidates are selected using the criteria described in Section~\ref{Sec:cluster}, and classified as class\,I, flat-spectrum, or class\,II objects by their spectral index estimated from four IRAC bands. The filled circles denote class\,I and flat-spectrum objects, open triangles class\,II objects. 
}\label{Fig:Glimpse_ccmap} 
\end{figure}

{\newrev We use the infrared color-color diagrams [3.6]-[4.5] vs. [4.5]-[8.0] and [3.6]-[4.5] vs. [5.8]-[8.0] to select candidate young  disk-bearing stars.} 
The selection criteria are as follows. In the [3.6]-[4.5] vs. [4.5]-[8.0] color-color diagram, objects are marked as YSOs if they meet the following criteria \citep[following][]{2007ApJ...669..327S}:  
{\newrev 
\begin{itemize} 
\item[] (1) [3.6]-[4.5]$>$0.6$\times$([4.5]-[8.0])-1.0,  
\item[] (2) [4.5]-[8.0]$<$2.8,  
\item[] (3) [3.6]-[4.5]$<$0.6$\times$([4.5]-[8.0])+0.3, 
\item[] (4) [3.6]-[4.5]$>$-([4.5]-[8.0])+0.85.  
\end{itemize} 
} 
In the [3.6]-[4.5] vs. [5.8]-[8.0] color-color diagram, YSOs must obey the criteria:  
{\newrev 
\begin{itemize} 
\item[] (1) [3.6]-[4.5] $\ge$0, and [5.8]-[8.0] $\ge$0.4 \citep{2004ApJS..154..363A},  
\item[] (2) [3.6]-[4.5] $\ge$ 0.67-([5.8]-[8.0])$\times$0.67, 
\end{itemize} 
} 
 where the latter criterion serves to remove the contamination arising from uncertainties in the IRAC photometry. Finally, we clean the thus constructed YSO catalog from contamination by AGNs and galaxies{, \newrev and exclude 4 objects}, according to the criteria of \cite{2008ApJ...674..336G}. Based on their IRAC spectral index we divide the YSOs into class\,I/flat, and class\,II types \citep{1987IAUS..115....1L}. In \fig~\ref{Fig:Glimpse_ccmap}, we show IRAC color-color diagrams of all detected objects, indicating the color boundaries used for the selection of YSOs and showing the class~I/flat and class~II sources as well as the objects not marked as disk-bearing YSOs with different colors.  In total we identify 64 class\,I/Flat sources and 244 class\,II sources in the field of Fig.~\ref{Fig:mm_map}.

In \fig~\ref{Fig:mm_map}, we show the spatial distribution of the objects identified as young stars with disks in NGC\,6357, to which we will in this subsection simply refer as ``YSOs'', reminding the reader that the population of young stars \emph{without} disks that is almost certainly also present is not included here. Inside the three bubbles CS\,61, CS\,59, and CS\,63 the number density of YSOs is obviously enhanced, suggesting recent star formation in these regions. In most star-forming regions a strong positional coincidence between young class\,I/flat YSOs and dense molecular cores is observed \citep[e.g.][]{2009ApJS..181..321E,2009A&A...504..461F}. However, in the NGC\,6357 complex this correlation is less obvious. The reason for this is currently unclear, possibly the parental molecular clouds of these YSOs have only recently been eroded by nearby massive stars, and star formation is still inactive in the molecular shells surrounding the bubbles.
 
A closer inspection of \fig~\ref{Fig:mm_map} reveals an interesting aspect of the spatial distribution of the class~I/flat sources: in the regions marked ``A'' and ``B'' they appear to form arcs subtending the bubble CS\,61. If we approximate the distribution of the sources in regions A and B with a circle we find that the center of the best-fitting circle lies very close to the center of bubble CS\,61. This is suggestive of some role of the massive stars in CS\,61 in (triggering) the presumably recent formation of the class~I/flat sources in regions A and B.  Similar configurations, of young stars forming an arc around an older stellar population, have been found in other star forming regions, e.g., Tr\,37,  RCW\,82, RCW\,120 \citep{2005AJ....130..188S,2009A&A...494..987P,2009A&A...496..177D}. 
 
\begin{figure} 
\centering 
\includegraphics[width=0.85\columnwidth]{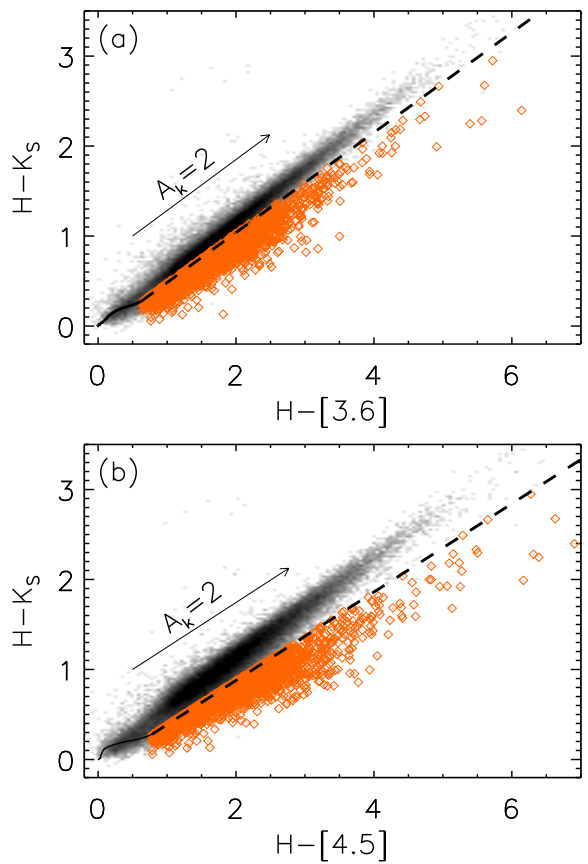} 
\caption{(a) $H-K_{\rm s}$ vs. $H-$[3.6] and (b) $H-K_{\rm s}$ vs. $H-$[4.5] color-color diagrams.  The solid lines show the intrinsic colors of diskless stars. {\rev The dashed lines and arrows present the extinction laws. The lengths of arrows show a extinction of 2\,mag at $K$ band.} The open diamonds are the YSO candidates which show excess emission in both [3.6] and [4.5] bands.}\label{Fig:Glimpse_2mass_ccmap} 
\end{figure}

The aforementioned criteria for YSO selection require that the stars have been detected in all four IRAC bands. This may cause many low-mass, low luminosity members to be missed due to the limited sensitivity and the strong nebular background in the IRAC [5.8] and [8.0] bands. To probe also the low-mass population, we resort to $H-K_{\rm s}$ vs. $H-$[3.6] vs. and  $H-K_{\rm s}$ vs. $H-$[4.5] color-color diagrams. In \fig\,\ref{Fig:Glimpse_2mass_ccmap}, we show two color-color diagrams on which the colors of unobscured diskless stars are indicated, together with the reddening vector \citep{2005ApJ...619..931I}. Stars without a clear infrared excess lie in a narrow band on the top-left side of, and roughly parallel to, the reddening vector. Some of these will be cluster members that have already lost their disks, or at least the hot inner parts thereof, but most will be unrelated foreground or background stars. The stars to the bottom-right side of the reddening vector have too red $H-$[3.6] and $H-$[4.5] colors to be explained by the reddening of diskless stars, indicating that they have substantial excess emission above the photospheric level at near-infrared wavelengths. Thus, we identify these sources as YSOs if they show excess emissions in both the IRAC [3.6] and [4.5] bands. In the following we will refer to them as ``lower-luminosity YSOs'', as opposed the brighter part of the sample that was detected in all four IRAC bands. This distinction, however, is set merely by observational detection limits and the respective stars likely form the fainter and brighter part of the same population.

\begin{figure*} 
\centering 
\includegraphics[width=11cm]{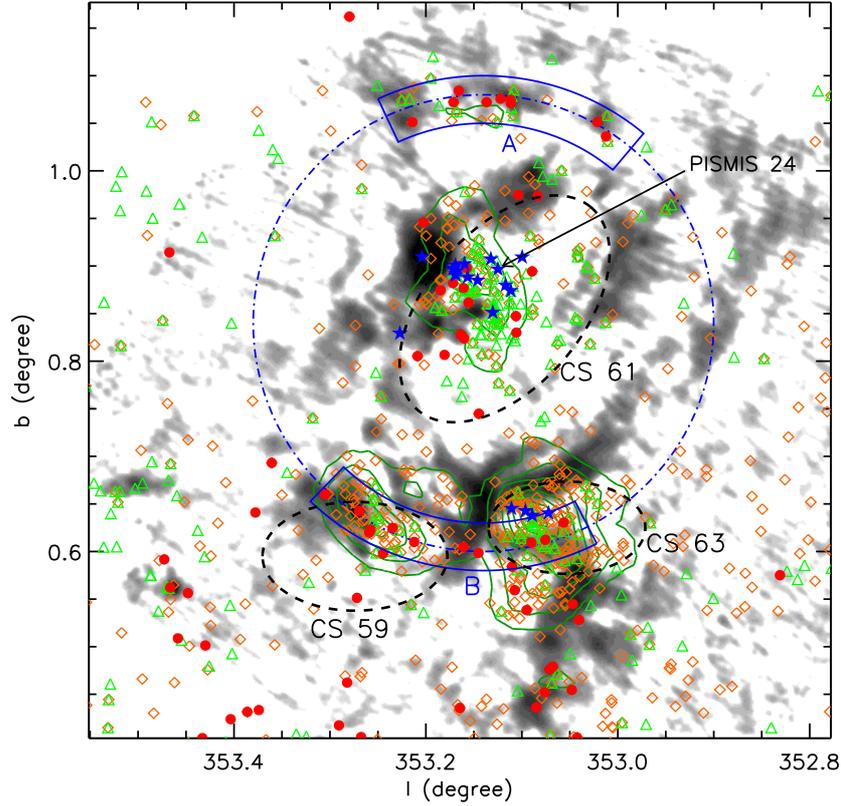} 
\caption{1.2\,mm emission map of NGC\,6357 complex. Filled circles mark the Flat/class\,I objects, open triangles mark the class\,II objects. The filled asterisks show known massive stars in Pismis\,24. The open diamonds show the YSO candidates selected from Fig~\ref{Fig:Glimpse_2mass_ccmap}. The surface density  of YSO candidates is denoted with contours. Three overdense regions are identified, corresponding to Pismis\,24  and two newly found clusters.}\label{Fig:cluster} 
\end{figure*}

In \fig\,\ref{Fig:cluster} we show the distribution of lower-luminosity YSOs selected as described in the previous paragraph and illustrated in \fig\,\ref{Fig:Glimpse_2mass_ccmap}, together with brighter class~I/flat and class~II objects that were detected in all IRAC bands. We plot the position of each of the lower-luminosity YSOs and also calculate their surface density, which we plot as contours in \fig\,\ref{Fig:cluster}. In the NGC\,6357 complex there are three regions where the distribution of the lower-luminosity YSOs shows an obvious overdensity, which coincide with the concentrations of the higher luminosity YSOs. All three coincide with the bubbles discussed earlier. The over-density of low-luminosity YSOs associated with bubble CS\,61 is the Pismis\,24 cluster, the other two are newly discovered young clusters that are spatially coincident with bubbles CS\,59 and CS\,63. We will refer to them as the ``CS\,59 and CS\,63 clusters'' hereafter. The spatial association of clusters and bubbles in NGC\,6357 can be understood since clusters form in dense parts of molecular clouds, whereafter the high UV flux and stellar winds of the massive cluster members create the bubbles. 
 
  Contrary to the central regions where we have X-ray and optical data in addition to the Spitzer photometry, the YSO identification in the largest part of the GLIMPSE field relies solely on color-color diagram analysis. This makes the sample more susceptible to contamination by galactic and extra-galactic sources unrelated to the cluster. NGC\,6357 is located in the galactic plane looking inward through the Milky Way ($l$=353\degree, $b$=$+$0.9\degree) and the galactic extinction in that direction is so high that most extra-galactic sources will be effectively absorbed even at 4.5\,\mum. Galactic contamination may arise from post-main sequence objects, mostly in the background, and young stars in foreground or background star-forming regions. The number density of the former sources is expected to be low and their distribution uniform, and thus the post-main sequence population should not substantially affect our results in a statistical sense. {\newrev The contamination from young stars is more difficult to evaluate. They would mostly be distributed in clusters or loose associations, and to remove them requires knowing their distances. The fact that many young objects are spatially coincident with high-density gas, and the young clusters of high-mass stars occupy holes in the gaseous distribution suggest that the YSOs and the surrounding gas are physically related. Furthermore,  \citet{2010A&A...515A..55R} have shown that the molecular gas spatially associated with cluster CS\,59 and CS\,63 has similar radial velocity to that of NGC\,6357, suggesting they are at the same distance.}

\subsection{The Pismis\,24 cluster} 
In this section we will zoom in on the Pismis\,24 cluster. First we will re-assess the distance to Pismis\,24. Then we will investigate the stellar properties of the cluster and derive the stellar masses and ages. Finally we will study the disk properties and frequency of the cluster members and compare them with those of other clusters.

\subsubsection{Distance to Pismis\,24} 
\label{sec:pis24_distance} 
\begin{figure*} 
\centering 
\includegraphics[width=13cm]{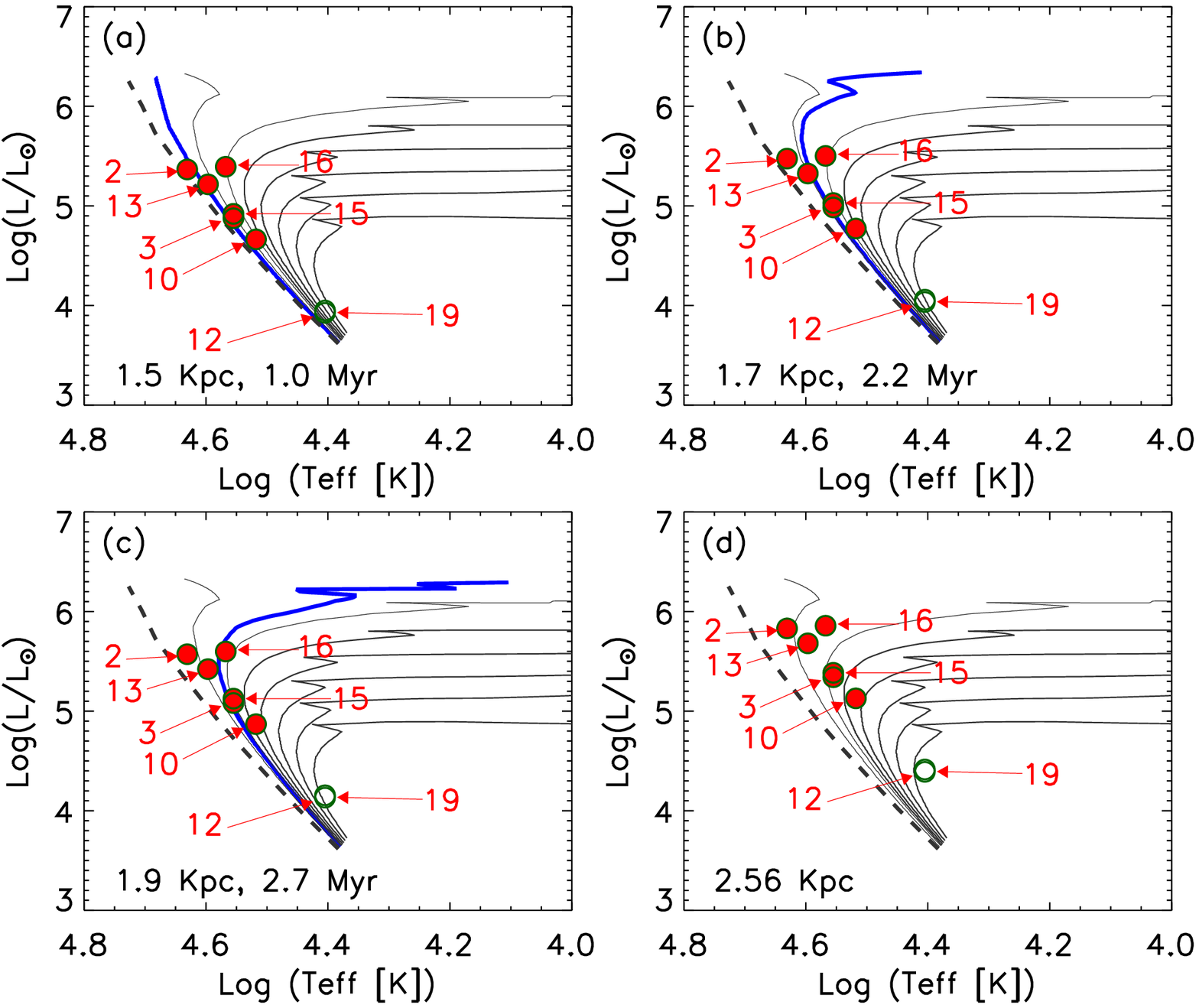} 
\caption{HR diagrams for six O-type main-sequence massive stars (Pis\,24-2, 13, 16, 3, 15, 10, filled circles), and two B-type main-sequence  stars (Pis\,24-12, 19, open circles) in Pismis\,24.  The massive stars  are marked with numbers (See Table~\ref{Tab:tab1}). The dashed lines represent the zero-age main sequence locus. The thin solid lines show the isochrones of  2, 3, 4, 5, 6, 8, and 10\,Myr. The thick solid lines present the best-fitting isochrones of 1, 2.2, and 2.7\,Myr in panels(a)(b)(c), respectively. The evolutionary tracks are from \citet{1992A&AS...96..269S}. The fitting gives a distance of 1.7$\pm$0.2\,kpc. {\newrev (d) HR diagram for massive stars with an assumed distance of 2.56\,kpc. There is no single isochrone that fits all six O-type main sequence stars.}}\label{Fig:HRD} 
\end{figure*}

There are some discussions in the literature about the distance of Pismis\,24, and published estimates range from 1.0 to 3.0~kpc. The most commonly adopted distance of 2.56\,kpc was derived by \citet{2001AJ....121.1050M} using observations of six massive stars, adopting their absolute magnitudes and intrinsic colors from the observed spectral types under the assumption {\rev that there is a unique absolute magnitude corresponding to each spectral type and luminosity class}, and matching these to the observed photometry. However, as shown in \fig\,\ref{Fig:HRD} the spectral types of massive stars do not uniquely constrain their absolute magnitudes without knowledge of their ages. In this work, we use the isochrone-fitting method to estimate the distance to Pismis\,24 and the age of the massive stars that are used as distance indicators. Therefore we need to put these stars on the HR diagram, which requires knowledge of the total luminosities and spectral types. We adopted the spectral types and optical photometry in the $B$ and $V$ bands from \citet{2001AJ....121.1050M} and complemented the optical data with photometry in the $J$, $H$, and $K_{\rm s}$ bands from the 2MASS catalog. We then performed SED fitting following the method described in \citet{2009A&A...504..461F}: we take a Kurucz model atmosphere spectrum with a fixed effective temperature corresponding to the observed spectral type, and fit a reddened and scaled version of this model spectrum to the observed photometry. In this fit we thus have only two free parameters: visual extinction $A_{\rm V}$ and the stellar angular diameter $\theta$. We adopt a standard extinction law \citep{1989ApJ...345..245C} with a total to selective extinction ratio of $R_{\rm V}$\,=\,3.1. We used the $BVJHK$ band photometry  to do the SED fitting. We calculated model fluxes by integrating the intensity of the (reddened) model atmospheres over the spectral response curve of the system for each filter. The synthetic photometry was then compared with the observations. By varying the parameters and minimizing the resulting $\chi^{2}$ we obtain the optimum values for extinction and the angular diameter of each star, from which we can derive the luminosities of the stars assuming a distance.

 We used the six O-type main-sequence stars (see Fig.~\ref{Fig:HRD}) to estimate the distance of Pismis\,24. By requiring that the six stars are located above the zero-age main sequence locus, we estimated a lower limit on the distance, which is $\sim$1.4\,kpc. We performed isochrone fitting to these six stars, assuming they formed coevally (see Fig.~\ref{Fig:HRD}). Using 1.4\,kpc as the lower limit of the distance, we varied the distance and age of the objects and found that the O-stars can be fitted by isochrones with  ages of $\sim$1-2.7\,Myr and distances of 1.7$\pm$0.2\,kpc (see Fig~\ref{Fig:HRD}). This puts NGC\,6357 at the same distance as the NGC\,6334 cloud \citep{2008hsf2.book..456P}, its direct neighbor on the sky, and suggest that both clouds are physically related instead of merely being close in projection.  Additional support for this is given by radio observations that show that the average radial velocity of NGC\,6357 is similar to that of NGC\,6334 (both $\sim-$4\,km s$^{-1}$), and that there are filamentary structures apparently connecting both complexes \citep{2010A&A...515A..55R} .  
 
{\newrev To test the influence on the resulting distance of a different $R_{\rm V}$ value, we varied $R_{\rm V}$ from 3.1 to 3.5 {\newnewrev \citep{2004AJ....127.2826B}}. We fitted the SEDs of the massive stars {\newnewrev using} the same method as described above, and derived their luminosities which are slightly higher ($\sim$9\%) than those estimated with $R_{\rm V}$=3.1. We also did  isochrone fitting to the six O-type main-sequence stars, which gave a slightly lower (5\%) distance  than that derived with $R_{\rm V}$=3.1. Therefore, we conclude that the distance estimate for Pismis 24 depends only very mildly on the assumed extinction law.}

\subsubsection{The heart of Pismis\,24} 
\label{sec:pis24_heart} 
 
There are 12 known massive stars in the Pismis\,24 cluster for which spectral type estimates exist. With the revised distance of 1.7\,kpc we estimate their total luminosities, masses, and foreground extinctions. The results are listed in \tab\,\ref{Tab:tab1}.
The median visual extinction of these stars is 5.8\,mag with a standard deviation of 0.5\,mag. Since these stars have likely already dissipated all their circumstellar material, the observed extinction should be entirely due to absorption by foreground dust. 
 
\addtocounter{table}{1} 
\begin{table*} 
\caption{Parameters for massive stars in Pismis\,24.}\label{Tab:tab1}           \label{table:1}   
\centering        
\begin{tabular}{lllcccc}  
\hline\hline              
 &RA & DEC &   &Lum\tablefootmark{b}&$A_{\rm v}$\tablefootmark{c}&Mass\tablefootmark{d} \\   
Name &(J2000) &(J2000)&Spt\tablefootmark{a}&($L_{\odot}$)&(mag)&(\Msun)\\ 

\hline                      
    Pis 24-1NE&17 24 43.497  &-34 11 56.86  &O3.5 If*     &5.89 &5.54&74 \\ 
    Pis 24-1SW&17 24 43.481  &-34 11 57.21  &O4 III       &5.81 &5.52&66 \\ 
    Pis 24-17 &17 24 44.73   &-34 12 02.7   &O3.5 III     &5.93 &6.34&78 \\ 
    Pis 24-2  &17 24 43.28   &-34 12 44.0   &O5.5 V(f)    &5.47 &5.83&43  \\ 
    Pis 24-13 &17 24 45.79   &-34 09 39.9   &O6.5 V((f))  &5.33 &6.39&35  \\ 
    Pis 24-16 &17 24 44.45   &-34 11 58.9   &O7.5 V       &5.50 &7.24&38 \\ 
    Pis 24-3  &17 24 42.30   &-34 13 21.3   &O8 V         &4.98 &5.82&25 \\ 
    Pis 24-15 &17 24 28.95   &-34 14 50.7   &O8 V         &5.03 &5.47&25 \\ 
    Pis 24-10 &17 24 36.04   &-34 14 00.5   &O9 V         &4.77 &5.80&20 \\ 
    Pis 24-18 &17 24 43.29   &-34 11 41.9   &B0.5 V       &4.47 &6.45&15 \\ 
    Pis 24-12 &17 24 42.27   &-34 11 41.2   &B1 V         &4.03 &5.58&11 \\ 
    Pis 24-19 &17 24 43.69   &-34 11 40.7   &B1 V         &4.06 &6.09&11 \\ 
\hline    
\end{tabular} 
\tablefoot{ 
\tablefoottext{a}{Spectral types from \citet{2001AJ....121.1050M} and \cite{2007ApJ...660.1480M}.} 
\tablefoottext{b}{The total luminosities for Pis 24-1NE and Pis 24-1SW are derived with the absolute $V$-band magnitudes using the bolometric corrections from \cite{1996ApJ...460..914V}, assuming a distance of 1.7\,kpc. The total luminosities for other stars are estimated from SED fitting (see \sek\,~\ref{sec:pis24_distance}).} 
\tablefoottext{c}{Besides Pis 24-1NE and Pis 24-1SW, the visual extinction for other stars come from SED fitting (see~\ref{sec:pis24_distance}). The visual extinction for Pis 24-1NE and Pis 24-1SW are from \cite{2007ApJ...660.1480M}.} 
\tablefoottext{d}{The stellar masses are derived using  the evolutionary tracks from \cite{1992A&AS...96..269S}.} 
} 
\end{table*} 
 
\begin{figure} 
\centering 
\includegraphics[width=\columnwidth]{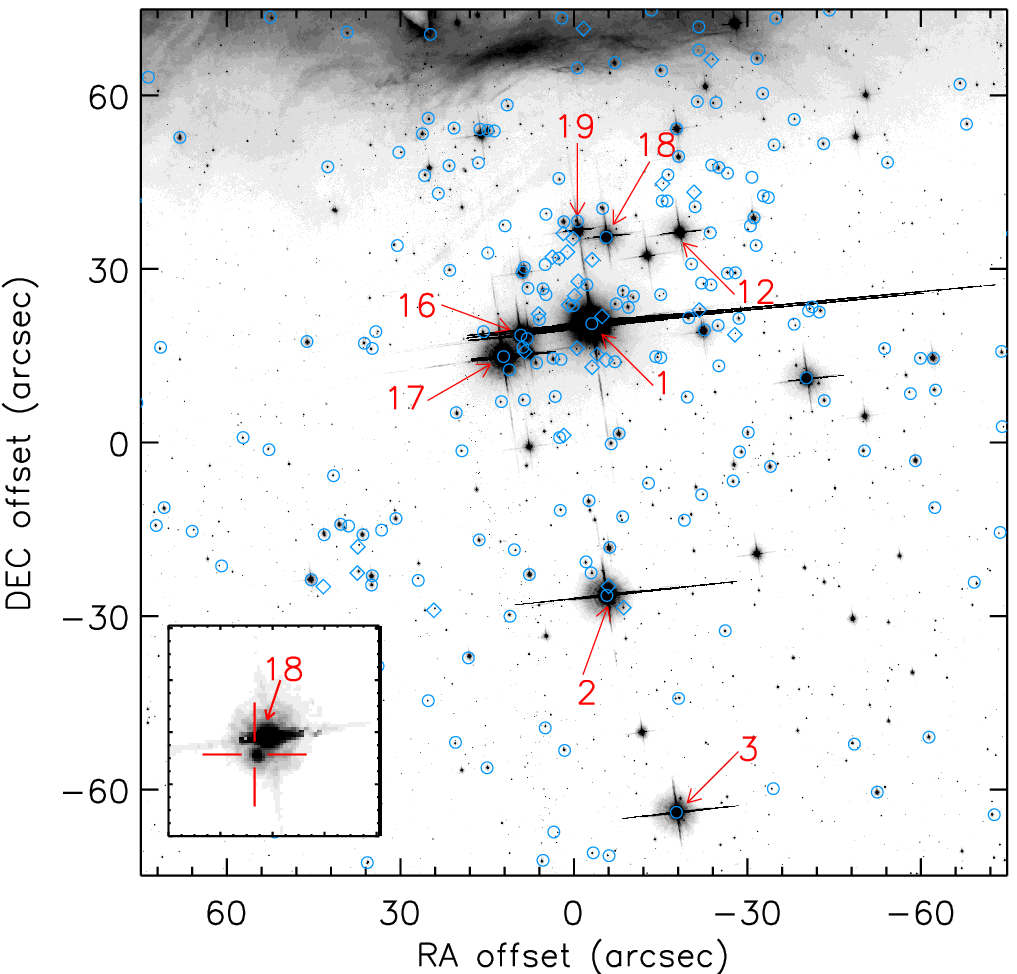} 
\caption{HST F850LP image of the center of Pismis\,24. The open circles mark the counterparts of X-ray sources previously identified in \citet{2007ApJS..168..100W}. The open diamonds represent counterparts of X-ray sources newly identified in the current work. The massive stars in this field are marked with numbers (See Table~\ref{Tab:tab1}). The inset shows the HST F550M image of Pis 24-18, which is resolved into a binary system. The plus sign marks the position of the X-ray emission source detected by \citet{2007ApJS..168..100W}.}\label{Fig:heart} 
\end{figure}

In \fig\,\ref{Fig:heart} we show an HST F850LP image of the center of Pismis\,24. In these high-resolution observations Pis\,24-18 is resolved into a binary system with a separation of 0\farcs45, corresponding to a projected distance of $\sim$765\,AU at 1.7\,kpc. \citet{2007ApJS..168..100W} detected X-ray emission from Pis\,24-18. We find that the secondary component of Pis\,24-18 matches the position of the detected X-ray source more closely, and therefore is the more likely counterpart of the X-ray source. In the whole field shown in \fig\,\ref{Fig:heart} there are 253 X-ray sources \citep{2007ApJS..168..100W}. Among these, 220 sources have detected optical or infrared counterparts. We used all detected X-ray sources \emph{with} optical/IR counterparts to calculate the surface density of the numbers of stars in the region, and find values of $\sim$800\,pc$^{-2}$ within a projected radius of 0.1\,pc from Pis\,24-1 and $\sim$350\,pc$^{-2}$ within 0.3\,pc from Pis\,24-1.  
 
\subsubsection{The low- and intermediate-mass population in Pismis\,24 }\label{Sec:spt_ext}

In this section we will inventory the stellar content of the Pismis\,24 star cluster in the low- and intermediate mass range. We will estimate the extinction distribution for the spectroscopy sample, for which we have reliable spectral types from our spectroscopic observations. We will assume that the cluster members {\newrev without} spectroscopy follow the same extinction distribution, in a statistical sense, as the ``spectroscopic'' sample. We will use an $R$~vs.~$R-I$ color-magnitude diagram, in which we de-redden all detected objects according to the derived extinction distribution, to estimate the mass and age distributions of all cluster members by comparison to the theoretical pre-main sequence tracks of \cite{2008ApJS..178...89D}.

\begin{figure} 
\centering 
\includegraphics[width=9cm]{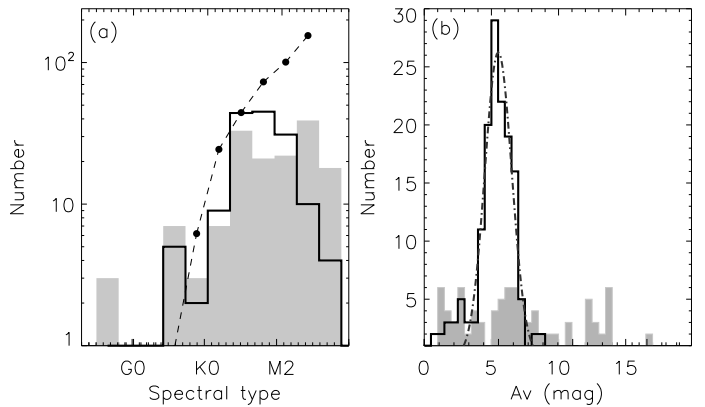} 
\caption{(a): the spectral-type distribution of our spectroscopic sample. The filled histogram {\newrev shows} the distribution of stars without X-ray emission. The open histograms display the distribution of X-ray emission stars. The dotted line connecting filled circles show the predicted distribution of spectral type for a 1\,Myr cluster with an IMF like that of the Trapezium cluster. The distribution is normalized to the number of K4-K6 stars. (b): the distribution of extinction for the spectroscopic sample. The filled histogram shows the distribution of stars without X-ray emission. The open histogram shows the distribution of X-ray emitting stars. The dash-dotted line denotes the Gaussian fit to the open histogram, peaking at $A_{\rm{V}}$=5.5~mag with a FWHM of 2.2~mag.}\label{Fig:dis_spt} 
\end{figure} 
 
\vspace{0.3cm} 
\noindent 
\noindent \textbf{(a) Spectral types and extinction}\\ 
\vspace{-0.3cm}

\noindent 
In total we have obtained the spectral types of 306 stars in the field of Pismis\,24. We have identified 151 of these as cluster members based on positional coincidence with a detected X-ray source. 155 stars in the spectroscopic sample were  {\newrev not} detected in the X-ray data, and were considered to be unrelated field objects.  We did not use the Li\,I $\lambda$6707 absorption line or H$\alpha$ emission line  as the indicator of youth properties due to the reasons: (1) the spectral resolution of our spectra is too low to detect the weak Li\,I $\lambda$6707 absorption line; (2) the spectra may suffer from H$\alpha$ line contamination from the surrounding nebula. 
In \fig\,\ref{Fig:dis_spt}(a) we show the spectral types of all stars in the spectroscopic sample. For the nonmembers the distribution is approximately flat between late K and late M spectral types. The distribution of the cluster members peaks around K5 and decreases towards later spectral types.  
In \fig\,\ref{Fig:dis_spt}(a) we also show a calculated distribution of spectral types for a model cluster with an age of 1\,Myr and a mass function as observed in the Trapezium cluster \citep{2002ApJ...573..366M}. Assuming that the Pismis\,24 cluster and Trapezium clusters have a similar IMF, we use the model spectral type distribution to evaluate the completeness level for the range of spectral types probed. As shown in \fig\,\ref{Fig:dis_spt}, the model distribution predicts many more M-type stars than are present in our spectroscopic sample, suggesting that the latter is substantially incomplete for the late spectral types. Assuming a foreground visual extinction of 6\,mag and a distance of 1.7\,kpc, the $R$-band  magnitude for a PMS {\newrev star} with a spectral type of M3 and an age of 1\,Myr is $\sim$23\,mag \citep{2008ApJS..178...89D}, which is indeed very faint for spectroscopic observations even with the VLT.

We used the observed $R-I$ colors and the intrinsic colors of the stars in the spectroscopic sample to estimate their extinctions, adopting the extinction law of \citet{1985ApJ...288..618R}. The intrinsic colors corresponding to each spectral type were taken from \citet{1998A&A...333..231B}. In  \fig\,\ref{Fig:dis_spt}(b) we show the resulting extinction distributions for the cluster members and the unrelated field stars. The two populations show very different extinction distributions. The unrelated field stars show a relatively flat distribution between $\sim$0 and $\sim$15 magnitudes, whereas the cluster members show a strongly peaked distribution centered around 5-6\,mag. The median extinction of the members of Pismis\,24 is $\sim$5.3\,mag, which is consistent with the extinction estimates of the known massive stars in the cluster (see \sek\,\ref{sec:pis24_heart}). Fitting the observed extinctions with a Gaussian distribution yields a peak value of 5.5\,mag and a FWHM of 2.3\,mag. In the following we will use the fitted Gaussian distribution of extinctions to derive the probability distribution of masses and ages for each cluster member for which we have only photometric observations but no spectroscopy.

\vspace{0.3cm} 
\noindent 
\noindent \textbf{(b) Mass and age of Pismis\,24}\\ 
\vspace{-0.3cm} 
 
\begin{figure*} 
\centering 
\includegraphics[width=18cm]{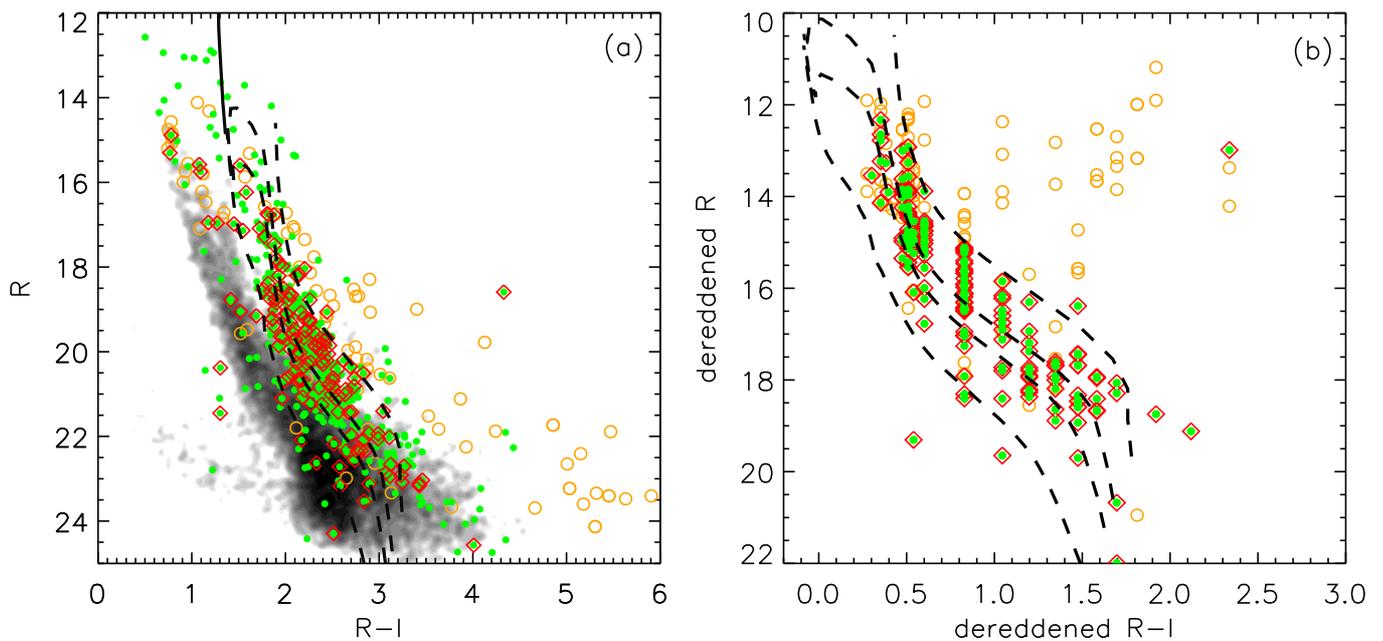} 
\caption{(a): $R$~vs~$R-I$ color-magnitude diagram for the stars detected on the VIMOS images. The dashed lines are PMS isochrones of  0.1, 1, 3, and 30\,Myr \citep{2008ApJS..178...89D}. The solid line represents the ZAMS \citep{1992A&AS...96..269S}. The isochrones are reddened by the mode of the extinction distribution of the spectroscopic members of Pismis\,24 cluster. The grey density map presents the distribution of all detected sources on the VIMOS images. The filled circles show the counterparts of X-ray sources. The open diamonds mark the X-ray emission stars that have been observed with VIMOS spectroscopy. The open circles present the stars without detected X-ray emissions that have been observed with VIMOS spectroscopy. (b): the dereddened $R$~vs~$R-I$ color-magnitude diagram for  the spectroscopic sample. The extinction for each star is estimated by comparing the observed $R-I$ color with the intrinsic $R-I$ color expected from the spectral type. The symbols are same as in panel (a).}\label{Fig:optical_ccmap} 
\end{figure*}

\noindent 
In \fig\,\ref{Fig:optical_ccmap}(a) we show the $R$~vs~$R-I$ color-magnitude diagram for all stars detected in the Pismis\,24 field. For comparison we also plot model isochrones of 0.1, 1, 3, and 30\,Myr \citep{2008ApJS..178...89D}. The isochrones have been reddened {\newrev by the visual extinction of 5.5\,mag (see~\sek\,\ref{Sec:spt_ext}(a))}. It can be noted that most of cluster members fall within the 0.1-3\,Myr isochrones.  In \fig\,\ref{Fig:optical_ccmap}(b), we show the dereddened $R$~vs.~$R-I$ color-magnitude diagram for our spectroscopic sample, along with the model isochrones of 0.1, 1, 3, and 30\,Myr \citep{2008ApJS..178...89D}.

We estimated the masses and ages of the spectroscopic members of Pismis\,24 from the dereddened $R$~vs.~$R-I$ color-magnitude diagram (see Fig~\ref{Fig:optical_ccmap}(b)) by comparison with the theoretical PMS evolutionary tracks \citep{2008ApJS..178...89D}. There are  several sets of pre-main sequence evolution tracks presented by  various authors \citep[e.g. ][]{1997MmSAI..68..807D,1998A&A...337..403B,1999ApJ...525..772P,2000A&A...358..593S,2008ApJS..178...89D}. In this work, we will adopt the  PMS evolution tracks from \citet{2008ApJS..178...89D}, as these have the best resolution in both mass and age. We stress, however, that there are substantial systematic differences between the different sets of tracks \citep[see ][for a detailed discussion of the various sets of PMS evolutionary tracks available in the literature]{2004ApJ...604..741H,2008ASPC..384..200H}, and our motives for choosing those by \citet{2008ApJS..178...89D} are pragmatic. Qualitatively, analyses such as ours do not depend on the specific set of tracks chosen, as long as the parameters of every object and every cluster are determined using the same set of theoretical tracks. 
 
\begin{figure*} 
\centering 
\includegraphics[width=17cm]{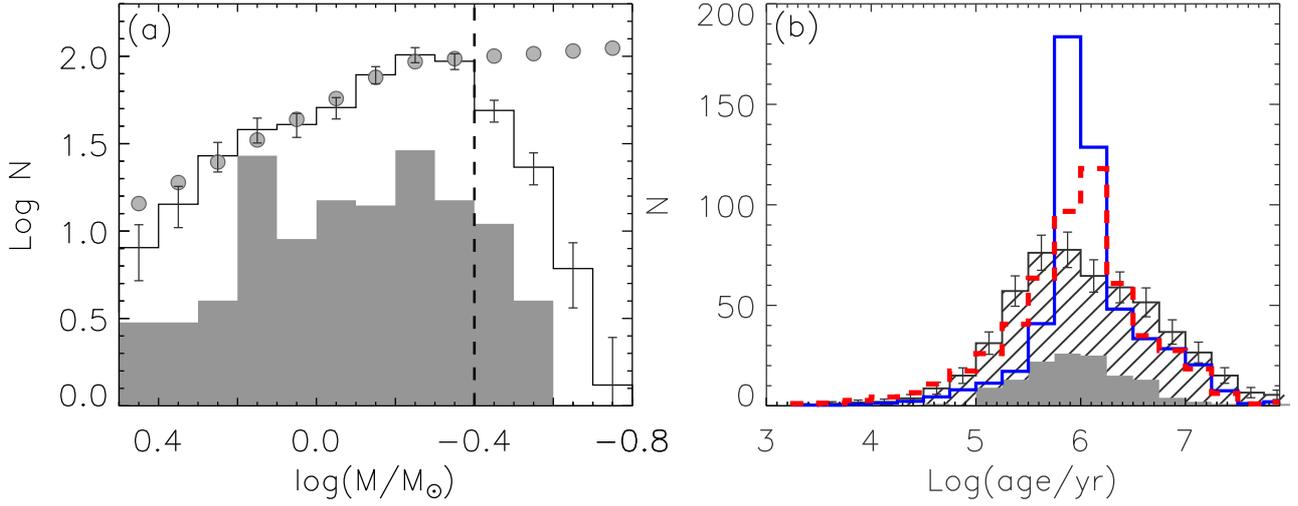} 
\caption{(a): The mass distribution of all X-ray emitting stars (open histogram), and X-ray emitting stars with well determined spectral types (filled histogram). The filled circles show the Trapezium IMF, scaled to match the open histogram. The dash line shows the completeness limit of the X-ray observations ($\sim$0.4\,\Msun). (b): The age distribution of all X-ray emitting stars (line-filled histograms) and X-ray sources with well determined spectral types (filled histograms). Both distributions yield a similar median age of $\sim$1\,Myr. The solid-line histograms show the coeval model. The dashed-line histograms represent the model with an age distribution of 0.9$\pm$0.9\,Myr}\label{Fig:dis_age} 
\end{figure*}

The results are listed in \tab\,\ref{tab:mass_age}. For the cluster members without spectroscopic observations we cannot accurately estimate the mass and age of any individual object since we do not know the {\newrev foreground} extinction towards individual stars. Instead, we derived a mass and age probability function for each object by drawing 2000 random samples from the extinction probability function derived from \fig\,\ref{Fig:dis_spt}(b). Thus we obtain 2000 ``virtual'' positions in the $R$~vs.~$R-I$ color-magnitude diagram for each observed star, each corresponding to a specific mass and age, and as a whole properly sampling the extinction distribution function {\newrev (see Appendix~\ref{Appen:model} for a detail description)}. The ensemble of the mass and age estimates of the cluster members yields a good representation of the actual cluster mass and age distributions, which are shown in Fig~\ref{Fig:dis_age}.  
 
. 
 
In \fig\,\ref{Fig:dis_age}(a) we compare the mass distribution of Pismis\,24 cluster with the IMF of the Trapezium cluster. The Trapezium IMF and the observed mass function of the Pismis\,24 cluster are very similar down to $\sim$0.4\,\Msun. At masses below 0.4\,\Msun \ both distributions are clearly different, with a substantial lack of observed stars in Pismis\,24 compared to the Trapezium cluster, which we attribute to incompleteness in the Pismis\,24 sample. {\newrev The average  10$\sigma$ detection limits in the VIMOS imaging data in the $R$ and $I$-bands  are 22.5 and 21.1\,mag, respectively. These limits vary from region to region due to the highly variable nebular background levels, in particular in the $R$-band.} From PMS evolutionary tracks we obtain apparent brightnesses of 22.3\,mag in $R$-band and 19.4 $I$-band, respectively, for a PMS star with a mass of 0.4\,\Msun, an age of 1\,Myr, and an extinction of $A_{\rm{V}}$=6\,mag \citep{2008ApJS..178...89D}. This confirms that the relative lack of Pismis\,24 members with masses below 0.4\,\Msun \ is at least to a large extent an observational bias due to the limited sensitivity of our VIMOS images.

{\newrev The mass distribution of the spectroscopic members is much flatter than that of all known cluster members in the mass range between $\sim$1.5\,\Msun \ and $\sim$0.4\,\Msun (see~\fig\,\ref{Fig:dis_age}(a)), again indicating that a large fraction of the low mass population has not been included in the spectroscopic data.  In addition the age distribution of the spectroscopic members and all known members of the Pismis 24 cluster (see~\fig\,\ref{Fig:dis_age}(b)) look very similar, though the former sample is very incomplete.} From these distributions, we estimate the median age of Pismis\,24. The spectroscopic members give a median age of 0.9\,Myr, and all known members give a median age of 1.0\,Myr.

 As shown in \fig\,\ref{Fig:dis_age}(b), the ages of the known members in Pismis\,24 show large spread. To investigate the statistical significance of the observed spread in age, we performed a simple Monte Carlo simulation. We first explored the possibility of coeval star formation in Pismis\,24. We used Monte Carlo techniques to generate a coeval population of 0.9 Myr old stars at a distance of 1.7\,kpc with a mass function of the Trapezium cluster. For each model star, we obtained the magnitude in $R$ and $I$ bands using the evolutionary tracks from \citet{2008ApJS..178...89D} according to the assumed mass and age. To mimic the observed photometric uncertainties, we varied the model photometry by adding random offsets drawn from a Gaussian distribution with a 1-sigma deviation of 0.1\,mag. To simulate the effect of extinction, we reddened the photometry with values drawn at random from the extinction probability function for Pismis\,24 (see \fig\,\ref{Fig:dis_spt}(b)), and included only synthetic stars with $R$-band magnitudes brighter than 23\,mag to well reproduce the observed color-magnitude diagram shown in Fig~\ref{Fig:optical_ccmap}(a). In one simulation, we produced a cluster with 1000 stars. We derived the age distribution from the model cluster with the same method that we applied for Pismis\,24. We have performed 10 simulations, and obtained an average age distribution from them. The resulting age distribution is shown in Fig~\ref{Fig:dis_age}(b). The age spread inferred from the observations is larger than what can be explained by an intrinsically coeval population with the aforementioned observational uncertainties. Therefore, we explore the possibility of an intrinsic  age spread for the members in Pismis\,24. We modified the original Monte Carlo simulation by randomly sampling the ages from a uniform distribution between 0-1.8\,Myr, and repeating the calculations in an otherwise identical way. {\newrev The resulting simulated age distribution (\fig\,\ref{Fig:dis_age}(b)) assuming an intrinsic age distribution matches the observations substantially better than a coeval population.} Still, the modeled peak in the age distribution around $\sim$1\,Myr is somewhat higher than the observed distribution, which may be due to other effects unconsidered in our simulations, e.g., stellar variability, unresolved binary, accretion activity including accretion history and current accretion rates, scattering effect from circumstellar disks \citep[e.g. ][]{2005MNRAS.363.1389B,2009ApJ...702L..27B,2010arXiv1006.3813G}. The stellar variability and unresolved binarity can induce the scatter of the apparent luminosity, therefore inducing an apparent spread in ages. Episodic accretion histories has also been proposed to explain the observed spread in HR diagrams by \citet{2009ApJ...702L..27B}. They show that an evolution including short episodes of vigorous accretion followed by longer quiescent phase can reproduce the observed luminosity spread in HR diagrams at ages of a few Myr years in the very low-mass range. Besides the accretion history, current accretion activity can produce excess emission, rendering the colors of young stars bluer and increasing the observed luminosity \citep{2010ApJ...722.1092D}. The scattering from circumstellar disks also makes the optical colors of young stars bluer, therefore affecting the derivation of physical parameters, e.g., extinction, and luminosity \citep{2010arXiv1006.3813G}. All these effects introduce scatter in the distribution of young populations on the HR diagram. 
 
\begin{figure*} 
\centering 
\includegraphics[width=13cm]{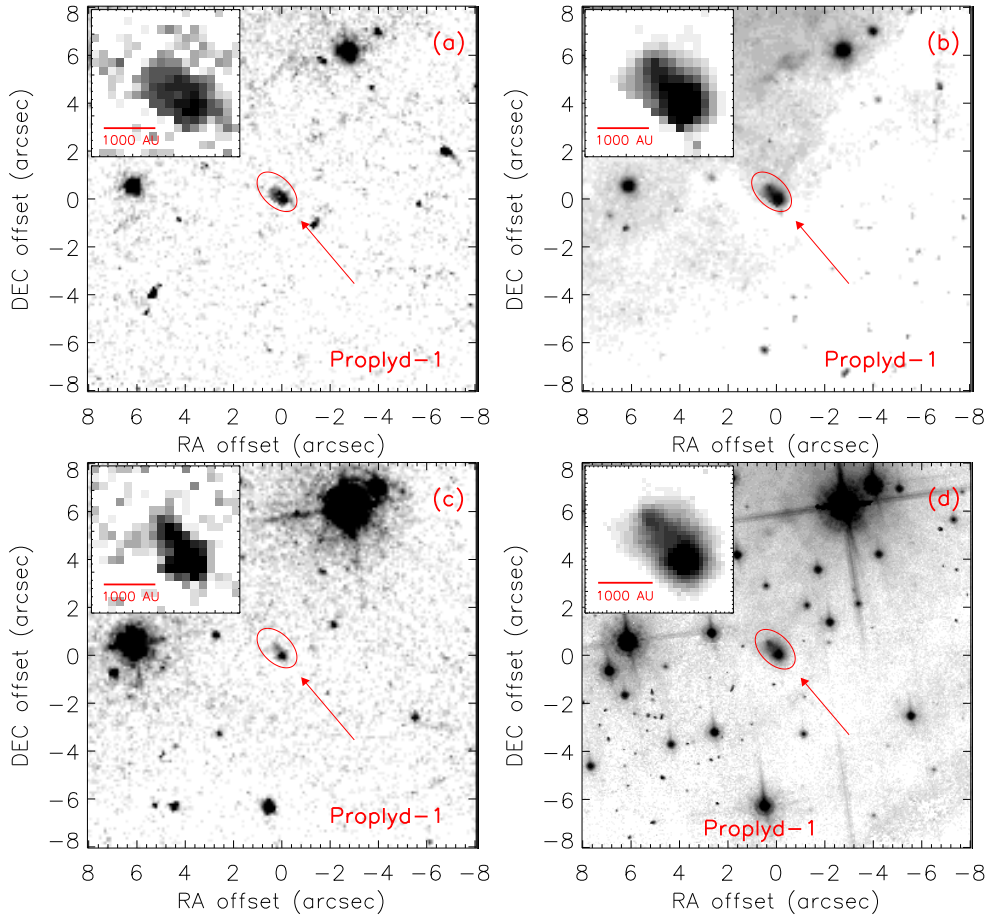} 
\caption{A photoevaporating disk candidate in Pismis\,24. Panels (a)(b)(c)(d) are centered on RA=17:24:45.26, DEC=-34:11:30.5 (J2000).  Panels (a)(b)(c) are HST images in the F502N, F656N, and F673N bands, which cover [O\,III], H$\alpha$ and [S\,II] respectively. Panel (d) shows the HST image in F850LP band. In each panel, the arrow shows the projected direction of Pis\,24-1 relative to the target. The inset in each panel shows a zoom-in of the proplyd in each HST band.}\label{Fig:hst} 
\end{figure*}

\subsubsection{Disk properties in Pismis\,24}\label{Sec:disk}

\vspace{0.3cm} 
\noindent 
\noindent \textbf{(a) Photoevaporating disks caught in the act}\\ 
\vspace{-0.3cm} 
 
\noindent 
{\newrev In \fig\,\ref{Fig:hst} we show an extended object  with the characteristic shape of proplyds \citep{2005ASPC..341..107H}.} These have been found in many star-forming regions harboring massive stars, e.g. the Orion Nebula \citep{1993ApJ...410..696O}, NGC\,3603 \citep{2000AJ....119..292B}, NGC\,2244 \citep{2006ApJ...650L..83B}, etc. The proplyds are interpreted as the outer disks of young stars that are being photoevaporated by EUV and FUV radiation from nearby massive stars. The energetic photons from neighboring massive stars heat gas in the outer disks to temperatures such that the sound speed exceeds the local escape velocity, allowing the gas to flow away \citep{2000prpl.conf..401H}. This scenario has been reproduced by simulations \citep[see e.g.][]{2000ApJ...539..258R}. The tail of proplyd-1 is pointing away from the most massive stellar system in the Pismis\,24 cluster, Pis\,24-1, suggesting that the latter is responsible for {\newrev its creation}. Proplyd-1 is located $\sim$0\farcs34 from Pis\,24-1, corresponding to a projected distance of 0.28\,pc.

In \fig\,\ref{Fig:hst}(a)(b)(c)(d) we show proplyd-1 in the HST F502N, F656N, F673N and 850LP bands, respectively. In all four bands proplyd-1 shows a bright, spatially extended  head, but its tail appears different in each band. In the F502N band the tail of proplyd-1 appears more diffuse and extended than in the F656N and F673N bands. The F502N, F656N, and F673N bands cover the [O\,III]5007, H$\alpha$, and [S\,II] 6717,6731 emission lines, respectively. The different appearance of the tails in the different bands can be attributed to {\newrev the} abundance of their agent. The H$\alpha$ and [S\,II]6717,6731 lines reach their maximum intensity at the hydrogen ionization front (H\,I-front), whereas the [O\,III]5007 line attains its maximum outside of the H\,I-front where EUV photons can still reach the oxygen and ionize O\,II \citep{2000ApJ...539..258R}. In the F850LP band proplyd-1 is firstly presented with a head-tail shape in high resolution ($\sim$0.1\arcsec). The tail shows a clear cone-like peak shape with extended diffuse emission, similar to the appearance in H$\alpha$. We estimate the length from the head to the tail of proplyd-1 in the F850LP band to be $\sim$1.\arcsec, corresponding to a projected length of $\sim$ 1700\,AU, comparable to the values found in simulations \citep{2000ApJ...539..258R}.

Proplyd-1 was also detected in our VIMOS images, with fluxes of 20.02$\pm$0.11\,mag in $R$ band, and 19.17$\pm$0.07\,mag in $I$ band. Its $R-I$ color of 0.85\,mag is unusually blue compared to that expected for young low-mass star in Pismis\,24: for a 1\,\Msun \ star with an age of 1\,Myr and behind 5.5\,mag of visual extinction we would expect an $R-I$ color of $\sim$2\,mag from PMS evolutionary models \citep{2008ApJS..178...89D}. The comparatively blue $R-I$ color of Proplyd-1 could be due to  (1) a dominant contribution of the H$\alpha$ emission line from the photoevaporating disk to the $R$-band flux; (2)the scattering effect by the evaporation circumstellar disks. Proplyd-1 is not present in the 2MASS catalog.  In the GLIMPSE catalog it is detected in the [3.6], [4.5], and [5.8] bands, with fluxes of 11.13$\pm$0.09\,mag, 10.54$\pm$0.31\,mag, and 9.30$\pm$0.10\,mag, respectively. These infrared fluxes indicate that proplyd-1 still has an optically thick inner disk. 
 
\begin{figure} 
\centering 
\includegraphics[width=\columnwidth]{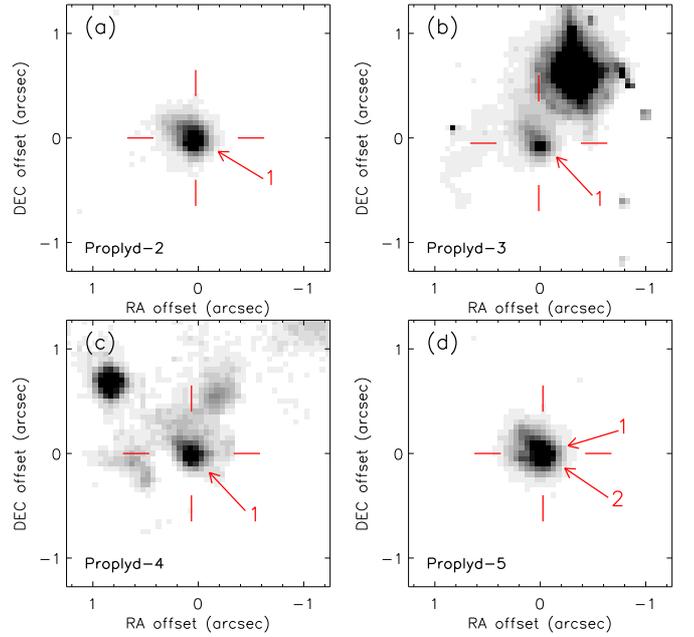} 
\caption{Four new photoevaporating disk candidates in Pismis\,24. Panels (a)(b)(c)(d) are the images from HST in the F850LP band, centered on RA=17:24:47.65, DEC=-34:11:25.4 (J2000), RA=17:24:45.88, DEC=-34:11:24.7 (J2000), RA=17:24:46.27, DEC=-34:11:19.7 (J2000), and RA=17:24:47.19, DEC=-34:12:10.8 (J2000), respectively. The numbers 1, and 2 on each panel are corresponding to  Pis\,24-1, and Pis\,24-2. The arrow on each panel shows the direction from Pis\,24-1 or Pis\,24-2 to each proplyd.}\label{Fig:hst2} 
\end{figure}

In addition to the previously known proplyd-1, we find four new proplyd candidates in the F850LP image, hereafter named Proplyd-2, Proplyd-3, Proplyd-4, and Proplyd-5, which are shown in \fig\,\ref{Fig:hst2}. Proplyd-2, 3, 4 show tails pointing away from Pis\,24-1, suggesting that the latter is responsible also for these proplyds. Careful inspection of proplyd-5 reveals two tails, a long one pointing away from Pis\,24-2 and a shorter tail pointing away from Pis\,24-1. This suggests that proplyd-5 is being photoevaporated by two neighboring massive stars simultaneously, with Pis\,24-2 dominating in this case. In our VIMOS images we have detected proplyd-2 and proplyd-5. Similar to proplyd-1, they show unusually blue $R-I$ colors (0.51\,mag and 0.28\,mag, respectively), confirming their proplyd nature. The projected distances from proplyds 2-5 to their ionizing stars are around 0.3-0.5\,pc, suggesting that massive stars like Pis\,24-1 and Pis\,24-2 can directly photoevaporate disks out to distances of at least 0.5\,pc. {\newrev Note that the five proplyds are too faint to be detected by the 2MASS survey and by Chandra. Therefore, we do not include them when estimating the disk frequency in the different distance bins from the massive stars in Pismis\,24 since we need to count the disk frequency at a uniform mass completeness level.}

\vspace{0.3cm} 
\noindent 
\noindent \textbf{(b) The disk frequency in Pismis\,24}\\ 
\vspace{-0.3cm} 
 
\begin{figure} 
\centering 
\includegraphics[width=\columnwidth]{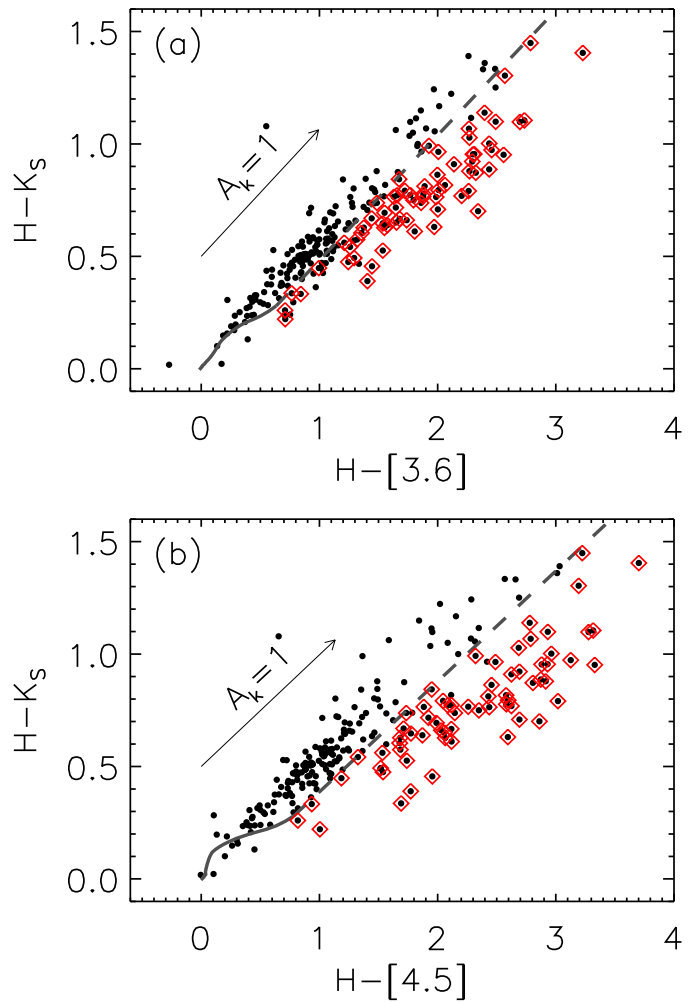} 
\caption{ (a) $H-K_{\rm s}$ vs. $H-$[3.6] and (b) $H-K_{\rm s}$ vs. $H-$[4.5] color-color diagrams for X-ray emitting stars in Pismis\,24 cluster. The filled circles represent all the members of Pismis\,24 cluster detected in $H$, $K_{\rm s}$, [3.6], and [4.5] bands.  The dashed lines and arrows represent the extinction laws. {\newrev The solid lines are the locus of colors for diskless stellar photospheres estimated from stellar model atmospheres \citep{2005ESASP.576..565B}.}  The arrow length denotes one magnitude of extinction in the K band. The open diamonds show the cluster members with inner disks which show excess emission in both [3.6] and [4.5] bands.}\label{Fig:deep_ccmap} 
\end{figure}

\noindent 
We will now analyze the ``disk frequency'' in the Pismis\,24 star cluster, i.e. which fraction of cluster members (still) show near-infrared excess emission indicative of optically thick material in the \emph{inner} disk regions. We will use the $H-K_{\rm s}$ vs. $H-$[3.6] and $H-K_{\rm s}$  vs. $H-$[4.5] color-color diagrams to identify stars with inner disks. These colors are more suitable than the classical IRAC [3.6]-[4.5] vs. [5.8]-[8.0] color-color diagram in the case of Pismis\,24, since requiring objects to be detected in \emph{all} IRAC bands causes the sample to be strongly biased towards relatively luminous sources and sources with disks. By including only the data from the two most sensitive IRAC bands we get a much more representative sample. Exclusion of the 5.8 and 8.0~\mum \ IRAC bands does cause potential transition disk objects, that show infrared excess only in the long wavelength IRAC bands, to be excluded from the disk {\newrev frequency} statistics. In \fig\,\ref{Fig:deep_ccmap} we show the colors of all 279 cluster members that are detected in  each of the $H$, $K_{\rm s}$, [3.6], and [4.5] bands. We identify stars as having inner disks if they show excess emission in both the IRAC [3.6] and [4.5] bands. In total, 83 cluster members harbor inner disks, yielding an inner disk frequency of 30$\pm$3\% in Pismis\,24.

\begin{figure} 
\centering 
\includegraphics[width=\columnwidth]{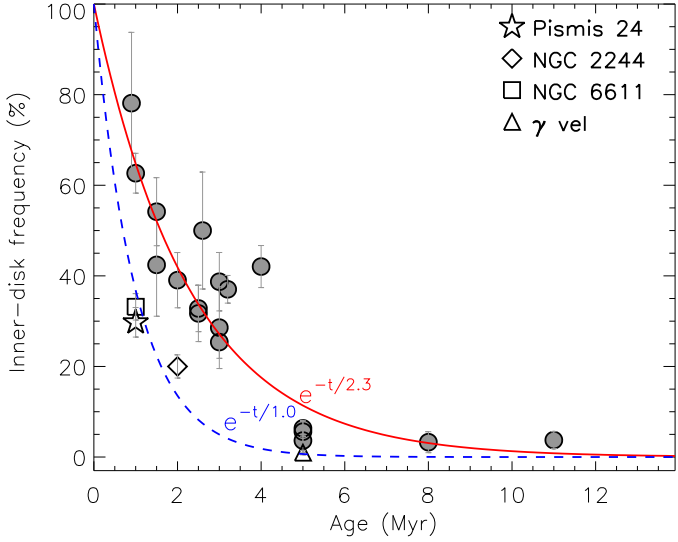} 
\caption{The inner disk frequencies (\fid) in different star formation regions as a function of age. The frequencies are estimated from  $H-K_{\rm s}$ vs. $K-$[3.6] and $H-K_{\rm s}$ vs. $K-$[4.5] color-color diagrams.  The frequencies are estimated for stars with masses larger than 0.5\,\Msun, which is the mass completeness limit for Pismis\,24 cluster. The dashed line  is the fit to inner disk frequencies of the four clusters, Pismis\,24, NGC\,2244, NGC\,6611, and $\gamma$\,vel, which gives \fid=$e^{-t/1.0}$, whereas the solid line is the fit to all other star formation regions, which is \fid\,=\,$e^{-t/2.3}$. Here t is the age in Myr.}\label{Fig:disk_frequency}  
\end{figure}

For comparison we estimate the inner-disk frequencies of YSOs in other star formation regions (SFRs) using the same diagnostic as employed for the Pismis\,24 cluster. A detailed description for each SFR is presented in Appendix~\ref{Appen:SFR}. In order to make a meaningful comparison we should calculate the disk frequency statistics for YSOs in the same mass range in each region since disk frequencies can depend on stellar mass \citep[e.g.][]{2006AJ....131.1574L,2008ApJ...675.1375L,2009A&A...504..461F,2007ApJ...662.1067H}, especially for YSO populations older than 3\,Myr \citep{2009ApJ...695.1210K}. The lowest mass that we use for selecting YSOs for our comparative disk frequency study is set by the mass completeness limit of the Pismis\,24 data. In the latter cluster the mass completeness is mainly limited by the 2MASS $H$ and $K_{\rm s}$ band data. Given a foreground extinction of $A_{\rm v}$$\sim$6\,mag, the 2MASS $H$ and $K_{\rm s}$ band magnitude limits of $\sim$15 and 14.3\,mag ({\rev 10$\sigma$}), and an age of 1\,Myr for Pismis\,24, the mass completeness for detection of the photospheric emission in the 2MASS data is approximately $\sim$0.5\,\Msun, and we will therefore consider only objects with masses above this limit in our comparative study. The resulting inner-disk frequencies (\fid) of each SFR as a function of their ages is shown in \fig\,\ref{Fig:disk_frequency}. For comparison we also plot the age-dependency of the \emph{accretor frequency}, which is also proxy for the presence of material in the inner disk, as obtained by \citet{2010A&A...510A..72F} by fitting an exponential profile to the observed accretor frequencies of a number of star forming regions. In most SFRs the inner disk frequency that we derive matches the accretor frequency behavior derived by \citet{2010A&A...510A..72F} very well. Four clusters, however, show substantially lower inner disk frequencies: Pismis\,24, NGC\,2244, NGC\,6611 and $\gamma$\,vel\,clusters. These four clusters all harbor extremely massive stars (see \tab\,\ref{tab:disk_frequency}).

\vspace{0.3cm} 
 
\begin{figure} 
\centering 
\includegraphics[width=\columnwidth]{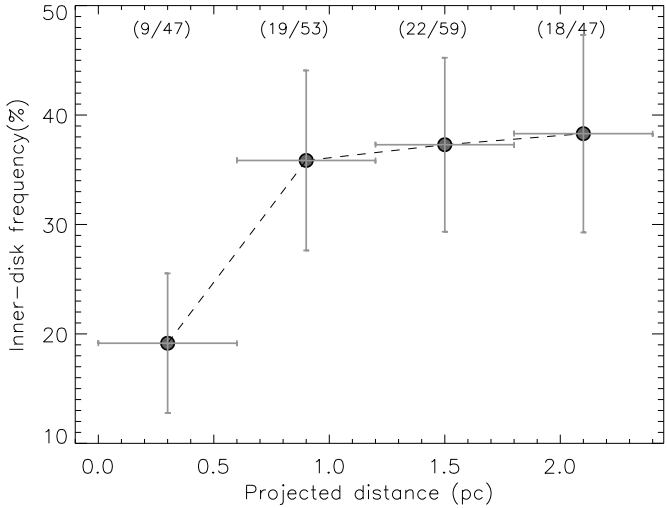} 
\caption{The inner disk frequency as a function of projected distance from Pismis 24-1, the most massive stellar system in the Pismis\,24 cluster.  Absolute number counts for each bin are given at the top of the panel.}\label{Fig:df_pos} 
\end{figure}

To see whether the inner disk frequency depends on location within the Pismis\,24 cluster we calculated it in four projected distance bins from Pis\,24-1 (the dominant UV-photon emitter): $\le$0.6\,pc, 0.6-1.2\,pc, 1.2-1.8\,pc, and 1.8-2.4\,pc. The result is shown in \fig\,\ref{Fig:df_pos}. In the innermost distance bin the disk frequency is $\sim$19\%. In the other three bins it is approximately constant at a substantially higher value of $\sim$36-38\%. The decrease of the disk frequency near Pis\,24-1 is a $\sim$2$\sigma$ effect in our data.  Decreased disk frequencies in the immediate vicinity of massive stars have been found in several massive clusters, e.g. NGC\,2244, NGC\,6611, and the Arches cluster \citep{2007ApJ...660.1532B,2009A&A...496..453G,2010arXiv1006.1004S}, suggesting rapid destruction of circumstellar disks in such environments.

\begin{figure} 
\centering 
\includegraphics[width=\columnwidth]{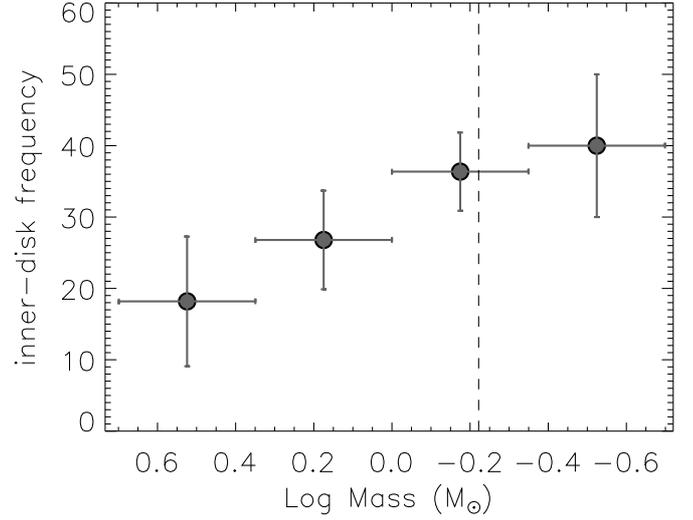} 
\caption{The inner disk frequency as a function of stellar mass. The dashed line marks the completeness limit of the sample.}\label{Fig:df_mass} 
\end{figure} 
 
In \fig\,\ref{Fig:df_mass}, we show the disk frequency as a function of stellar mass in the Pismis\,24 cluster. The adopted mass for each star is the median value of its mass probability function (see section~\ref{Sec:spt_ext}(b)). As shown in \fig\,\ref{Fig:df_mass}, the disk frequency increases with decreasing stellar mass in Pismis\,24. This is different from the behavior in some small clusters or isolated SFRs at similar ages of $\sim$1\,Myr \citep[e.g. NGC\,2068/2071, L1641, and Taurus, ][]{2009A&A...504..461F,2010ApJS..186..111L}, but similar to the one in some relative older clusters \citep[e.g. IC\,348, Tr\,37, and IC\,1795, ][]{2009ApJ...695.1210K, Veronica10}. Within Pismis\,24 we have compared the spatial distribution of cluster members of sub-solar mass with that of members of 1\,\Msun \ or more. The two distributions are not significantly different, and therefore we cannot attribute the lower disk frequency around YSOs with masses above 1\,\Msun \ to comparatively close proximity to the massive stars. Our observations suggest that the disks around $\sim$solar and intermediate mass stars evolve faster than those around lower mass stars in massive clusters like Pismis\,24.
 
\section{Discussion}\label{Sec:discussion}

\subsection{Disk evolution in Pismis\,24} 

As shown in \fig\,\ref{Fig:disk_frequency} the inner disk frequencies (\fid) in most SFRs follow the evolutionary function \fid\,=\,$e^{-t/2.3}$. Clusters in which extremely massive stars are present (Pismis\,24, NGC\,2244, NGC\,6611, and $\gamma$\,Vel), however, show a substantially quicker reduction of the inner disk frequency. For the later four clusters, fitting the same functional shape \fid\,=\,$e^{-t/\tau_0}$, we find $\tau_{0}$$\sim$1.0\,Myr. Therefore, the average inner disk lifetime in these massive cluster is only half of that in clusters without extremely massive stars. The disks in these massive clusters are then rapidly dissipated, but which physical mechanism is driving this development? A range of processes have been proposed to explain disk dissipation, including viscous evolution, gravitational interactions, and photoevaporation by the central star or neighboring massive stars. The latter two mechanisms, stellar encounters and photoevaporation, are potential explanations for the comparatively low disk frequency in massive clusters like Pismis\,24. 
 
\subsubsection{Stellar encounters} 
The stellar encounter mechanism involves a circumstellar disk and a nearby passing star. Mutual gravitational interaction can induce a significant loss of mass and angular momentum from the disk \citep{2006A&A...454..811P,2006ApJ...642.1140O,2007A&A...462..193P,2010A&A...509A..63O}. Simulations of stellar encounters in a cluster environment like the ONC have shown that stellar encounters are a potentially very important disk dissipation mechanism in the first several Myrs. When the stars involved in the encounters have unequal masses the disks are most affected. Therefore, in dense clusters like the ONC, the massive stars dominate the disk-mass loss \citep{2006ApJ...642.1140O}. The disks around these massive stars are also dissipated more quickly than those around intermediate- and low-mass stars \citep{2006A&A...454..811P}. Simulations show that, when the disk radii are scaled with the stellar mass, the disk frequencies decrease with increasing stellar masses due to stellar encounters in such cluster environments \citep{2006A&A...454..811P}. New calculations simulating a range of cluster environments show that the disk mass loss increases with the number density of stars in the cluster, but is not affected by the total number of cluster members \citep{2010A&A...509A..63O}. In less dense clusters, i.e. four times sparser than the ONC, the simulations still show substantial disk dissipation due to stellar encounters \citep{2010A&A...509A..63O}.  
 
We {\newrev estimated} the number density of stars in Pismis\,24 from the members selected from the Chandra observations. The X-ray data are complete down to $\sim$0.4\,\Msun, and the IMF of Pismis\,24 closely resembles a standard IMF as that of the Trapezium cluster in the mass range accessible to observations. In order to calculate the total number density of stars {\rev with masses larger than 0.08\,\Msun\citep[the minimum mass used in the simulations of ][]{2010A&A...509A..63O}}, we assumed that the IMF of Pismis\,24 follows the standard curve also at masses below 0.4\,\Msun, where we do not have proper observational constraints. We furthermore assumed the cluster members to be spherically distributed. In this way we derived the number density of stars to be approximately 1.5$\times$10$^{4}$\,pc$^{-3}$ within 0.1\,pc around Pis\,24-1, and 2.1$\times$10$^{3}$\,pc$^{-3}$ within 0.3\,pc around Pis\,24-1, somewhat lower than those in ONC, where the number density of stars to be approximately 2.4$\times$10$^{4}$\,pc$^{-3}$ within 0.1\,pc around the ONC center, and 3.1$\times$10$^{3}$\,pc$^{-3}$ within 0.3\,pc around the ONC center. Simulations predict that the disk frequency is reduced to approximately 85\% of the assumed initial disk frequency of 100\% due to the stellar encounters in clusters with such number densities of stars \citep{2010A&A...509A..63O}. Even though this decrease is substantial, it falls short of explaining the observed inner disk frequency within 0.6\,pc around Pis\,24-1, which we have shown to be only 19$\pm$6\%. Therefore, stellar encounters alone can hardly account for the quick dissipation of disks in the Pismis\,24 cluster. 
 
\subsubsection{Photoevaporation} 
In a cluster environment like Pismis\,24, where dozens of massive stars are present, photoevaporation is another mechanism to dissipate the circumstellar disks around the intermediate- and low-mass cluster members. UV photons from massive stars can heat the gas in the disk surface to temperatures at which the sound speed of the gas exceeds the velocity needed to escape the gravitational potential well of the star, inducing a gas flow away from the disk \citep{1998ApJ...499..758J,2000prpl.conf..401H}. In photoevaporation models a neutral gas flow from the disk surface is driven by non-ionizing FUV photons. EUV photons do not reach the disk surface because they are absorbed by the outflowing material. Further from the disk, where the densities are lower, EUV photons can penetrate and ionize the outflowing gas, forming an ionization front \citep{1998ApJ...499..758J,1999ApJ...515..669S,2000ApJ...539..258R}.  
 
The mass-loss rates from photoevaporating disks depend on a number of parameters, in particular on the intensity of the impinging UV field (especially the FUV flux), the disk radii, and the stellar masses. Model calculations predict the mass loss rates from disks due to photoevaporation in the Trapezium cluster to be on the order of 10$^{-7}$\,\Msunyr \ up to a projected distance of 0.2\,pc away from the ionizing massive star $\theta^{1}$C\,Ori, which has a spectral type of O6 \citep{1998ApJ...499..758J,1999ApJ...515..669S,1998AJ....116..322H}. Spectroscopic observations have confirmed these mass-loss rates \citep{1999AJ....118.2350H}. The photoevaporation process effectively dissipates circumstellar disks from the outside inward, up to the gravitation radius ($r_{\rm g}$), where the escape velocity equals the sound speed, which in turn is determined by the UV heating \citep{2000prpl.conf..401H}. For a star with a mass of 0.5\,\Msun \ the gravitation radius is estimated to be $\sim$60\,AU, assuming that the disk surface is heated to 1000\,K \citep{2004ApJ...611..360A}. \citet{2004ApJ...611..360A} show that when the disk radius is less than $r_{\rm g}$ the mass loss rates are still considerable down to 0.2\,$r_{\rm g}$.  
 
In the Pismis\,24 cluster there are tens of massive stars and the total FUV photons luminosity is estimated to be $\sim$10 times that of Trapezium cluster. Therefore we expect photoevaporation to be very effective in dissipating the outer disks, down to several tens of AU, around young stars within a distance of 0.6\,pc from the center of Pismis\,24. The timescale for this dissipation could be less than 0.5\,Myr, assuming an initial disk mass of 0.05\,\Msun\, and mass-loss rates similar to those observed in the Trapezium cluster.   
 
\citet{2004ApJ...611..360A} predict that the photoevaporation-induced mass-loss rates become very low at disk radii below 0.2\,$r_{\rm g}$. It is almost impossible to photoevaporate disks down to radii $r\le$0.1$r_{\rm g}$ if the heating of gas in the disk surface is dominated by FUV photons that heat the gas to $\sim$ 1000\,K \citep{2004ASPC..323....3H}. When also EUV photons can reach the disk surface the gas temperature can reach $\sim$10$^{4}$\,K. If recombination dominates the opacity to EUV photons the mass-loss rate is given by the following approximate formula: 
\begin{equation} 
  \dot{M}\approx9\times10^{-8}(\frac{\Phi}{10^{49}\,\rm{s}^{-1}})^{1/2}(\frac{d}{10^{17}\rm{cm}})^{-1}(\frac{r_{\rm d}}{30 \rm{AU}})^{3/2}\, \ $\Msunyr$ 
\end{equation} 
where $\Phi$ is the EUV photon flux, $d$ is the distance of the circumstellar disk from the ionizing massive star, and $r_{\rm d}$ is the disk radius \citep{1998ApJ...499..758J,2010ARA&A..48...47A}. In the Pismis\,24 cluster the EUV fluxes are dominated by the triple system Pis\,24-1, from which the EUV photon flux is estimated to be $\sim$10$^{50}$ photons per second. For a circumstellar disk with an initial mass of 0.05\,\Msun\, an outer radius of 100\,AU, and a surface density profile $\Sigma\approx r^{\rm -3/2}$, there will be $\sim$3.5$\times$10$^{-3}$\,\Msun \ of material within 1\,AU, where the hot dust causing the excess emission in the IRAC [3.6] and [4.5] bands resides. To photoevaporate this amount of material at sub-AU radii within a time span of 1\,Myr requires a mass-loss rate of $\sim$3.5$\times$10$^{-9}$\,\Msunyr induced by EUV photons, which is available only very close to Pis\,24-1, within 0.02\,pc. Thus, the rapid dissipation of inner disks at radii below $\sim$1\,AU around Pismis\,24-1 cannot be explained by the direct effect of photoevaporation alone, since most of the cluster members are at distances $>>$ 0.02\,pc. Additional physics is required to explain the low inner disk frequency in the Pismis\,24 cluster.  
 
\citet{2007MNRAS.376.1350C} couples the mass-loss profiles from \citet{2004ApJ...611..360A} with a viscous disk evolution model. In her model, the inner parts of the circumstellar disk are continuously being accreted onto the central star while the outer part is dissipated by the photoevaporation due to UV irradiation from nearby massive stars. Once the outer parts of disk have been stripped of their gas in the photoevaporative flow, the inner disk can no longer be replenished by material from larger radii. The inner disk is then quickly drained by viscous accretion onto the central star. Adopting an accretion rate of 2.5$\times$10$^{-8}$\,\Msunyr, the median accretion rate for $\sim$solar mass T-Tauri stars in the Taurus star-forming region \citep{2007MNRAS.378..369N}, viscous accretion can drain the inner disk within a radius of 30\,AU within 1\,Myr, given an initial circumstellar disk mass of 0.05\,\Msun, a disk radius of 100\,AU, and a surface density profile $\Sigma\approx r^{\rm -3/2}$. Therefore, the combination of viscous accretion and photoevaporation can effectively dissipate the circumstellar disks of young stars within a radius of $\sim$0.6\,pc from the center of Pismis\,24 cluster. The dissipation timescale can be less than 1\,Myr, depending on the initial disk masses. 
 
The mass-loss rates from photoevaporation are less sensitive to the intensity of the FUV field than on the disk radius. When the FUV fluxes decrease by a factor of ten the mass-loss rates are reduced only by a factor of 2-3 \citep{2004ApJ...611..360A}. Therefore the mass-loss rate from a circumstellar disk induced by photoevaporation can still be 3-5$\times$10$^{-8}$\,\Msunyr \ at a distance of 2\,pc from the massive stars. This is still a substantial mass loss rate and a disk with a mass of 0.05\,\Msun \ can be dissipated on a timescale of $\sim$1\,Myr. For disks that are located at substantially larger distances from the massive stars, direct photoevaporation is likely of minor importance. Still, the radiation from the massive stars substantially increases the local EUV field and can contribute importantly to the ionization of material in the disk surface layers. This may in turn strongly increase the viscosity of the disk material through the magneto-rotational instability {\newnewrev (MRI)}, {\rev of which occurrence and strength depends on the ionization fraction inside the disk \citep{2000prpl.conf..589S}}, and lead to increased accretion rates and reduced disk dissipation timescales. 
 
\subsection{Hot inner disk evolution and transition disks}

The inner disk frequencies (\fid) in clusters that do not harbor extremely massive stars (see \sek\,\ref{Sec:disk}(b) and \fig\,\ref{Fig:disk_frequency}) follow the approximate age dependency \fid\,=\,$e^{-t/2.3}$, where $t$ is the cluster age in Myr. The inner disk frequencies are estimated using $H-K_{\rm s}$ vs. $H-$[3.6], and $H-K_{\rm s}$ vs. $H-$[4.5] color-color diagrams, and thus sources that show infrared excess \emph{only} at wavelengths longer than $\sim$5\,\mum\ are considered to be (inner-) ``diskless'' in the \fid \ statistic. This means that a substantial fraction of the so-called ``transition disks'' (TD) are not included in the numerator of the \fid \ statistic as we apply it. If one would include all four IRAC bands in calculating the disk frequency\footnote{In Pismis\,24 this would bias the sample towards comparatively luminous sources or sources with optically thick disks due to the limited sensitivity in the IRAC 5.8 and 8.0~\mum \ bands, and hence we use data from the more sensitive 3.6 and 4.5~\mum \ bands only. In more nearby clusters the long wavelength IRAC channels are sufficiently sensitive to detect ``naked'' photospheres down to substantially lower stellar masses, and meaningful disk frequencies can be calculated using data from all four IRAC bands.}, sources without excess emission in the 3.6 and 4.5~\mum \ IRAC channels but with infrared excess in the 5.8 or 8.0~\mum \ IRAC band would be counted as having a disk. This approach was chosen by \citet{2010A&A...510A..72F}, who find a decay of the disk frequency with cluster age according to $f_{\rm disk}$\,=\,$e^{-t/3.0}$. Note that here we distinguish between the ``inner disk frequency'' \fid, as traced by the data at $\lambda$\,$\lesssim$\,5\,\mum, and the ``disk frequency'' $f_{\rm disk}$ as traced by observations at $\lambda$\,$\lesssim$\,9\,\mum\footnote{In the current discussion we adopt the latter statistic as the \emph{total} disk frequency, thereby ignoring the fact that some objects exist that have no infrared excess at $\lambda$\,$\leq$\,9\,\mum \ but do show excess emission at e.g. 24\,\mum. Therefore, the $f_{\rm disk}$ statistic gives only a lower limit to the actual frequency of circumstellar disks. This approach is brought about by practical limitations: the availability and sensitivity of longer wavelength (MIPS) data is substantially less favorable than for the IRAC data. Also crowding/confusion is more problematic at long wavelengths, in particular for distant clusters like Pismis\,24}. Thus, the disk frequency \fid \ as traced only by the IRAC [3.6] and [4.5] bands decreases more quickly than $f_{\rm disk}$ as calculated including also the [5.8] and [8.0] channels. The difference can be attributed to the presence of transition disks, and from the above-mentioned relations we can easily calculate which fraction of the disk population is made up by the transition disks: 
%
$ 
f_{\rm TD} = \frac{e^{-t/3.0} - e^{-t/2.3}}{e^{-t/3.0}} 
$ 
, where $t$ is in units of Myr. In the following we will compare this function to $f_{\rm TD}$ values given in the literature. 
 
\begin{figure} 
\centering 
\includegraphics[width=8cm]{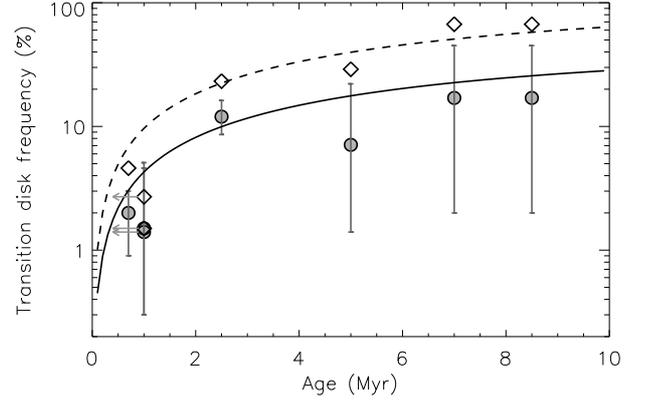} 
\caption{The frequencies of TDs (diamonds) and classical TDs (filled circles) among the disk population in different clusters/aggregates, including NGC\,1333, L1688, NGC\,2068/2071, IC\,348, OB1b, $\eta$\,Cha, OB1a/25\,Ori. The observational data are obtained from \citet{2010ApJ...708.1107M}, besides the age of NGC\,2068/2071 which is from \citet{2009A&A...504..461F}. The left-pointing arrows indicate that the ages for NGC\,1333, and L1688 are up-limits. The TD sample include classical TDs and weak excess TDs from \citet{2010ApJ...708.1107M}. The dashed line is the predicted frequency ($f_{\rm TD}=\frac{e^{-t/3.0} - e^{-t/2.3}}{e^{-t/3.0}}$) of TDs  among all the disk population as a function of age. The solid line is the predicted frequency ($f_{\rm CTD}$\,$=$\,0.44\,$f_{\rm TD}$) of classical TDs among all the disk population as a function of age.}\label{Fig:TD} 
\end{figure}

\citet{2010ApJ...708.1107M} use spectral indices between photometry in various IRAC and MIPS bands to select TDs in clusters and aggregates in the NGC\,1333, L1688, NGC\,2068/2071, IC\,348,{\rev Orion OB1a/25\,Ori and OB1b}, and $\eta$\,Cha. They use excess emission in the IRAC 5.8\,\mum \ and MIPS 24\mum \ bands to divide the TDs into three subclasses: classical TDs, weak excess TDs, and warm excess TDs. The classical TDs show little or no excess emission in [5.8] band and strong excess emissions in the MIPS [24] band. Weak excess TDs are similar to classical TDs in the IRAC [5.8] band, but are comparatively faint in the MIPS [24] band. The warm excess TDs show obvious excess emission in the IRAC [5.8] band. When comparing the fraction of the disk population that is considered to be a transition disk ($f_{\rm TD}$) we only count the classical and weak excess TDs in the sample of \citet{2010ApJ...708.1107M}, because warm excess TDs usually show substantial excess emission in the short wavelength IRAC bands and are thus not counted as transition disks in our $f_{\rm TD}$ statistic. In \fig\,\ref{Fig:TD} we plot our derived $f_{\rm TD}$ and compare it with the TD fractions in different SFRs from \citet{2010ApJ...708.1107M}. We find good agreement between $f_{\rm TD}(t)$ curve that we derive from the comparison of \fid$(t)$ \ and $f_{\rm disk}(t)$ and the more direct determination of $f_{\rm TD}(t)$ by \citet{2010ApJ...708.1107M}. 
 
Among the whole TD sample of \citet{2010ApJ...708.1107M} there are 44\% classical TDs. By scaling our $f_{\rm TD}$ function by a factor of 0.44, we obtain the fraction of the total disk population constituted by classical transition disks: $f_{\rm CTD}$\,$=$\,0.44\,$f_{\rm TD}$. In \fig\,\ref{Fig:TD} we also show $f_{\rm CTD}$, which well matches the $f_{\rm CTD}$ observed by \citet{2010ApJ...708.1107M}. 
 
\section{Summary} 
\label{Sec:summary} 
 
We have performed a detailed observational study of the massive star-forming region NGC\,6357, with a special focus on the central cluster Pismis\,24. Our study includes X-ray data from the literature, optical photometry and spectroscopy performed with VLT/VIMOS, HST archival data, infrared data from the literature as well as our own Spitzer imaging of the central cluster, and millimeter data. 
 
\vspace{0.2cm} 
 
Using infrared color-color diagrams we have searched for disk-bearing YSOs in the whole NGC\,6357 complex. These are concentrated in three clusters: the known Pismis\,24 cluster and two newly discovered clusters that are associated with the known bubbles CS\,59 and CS\,63. 
 
We have re-assessed the distance to Pismis\,24 by fitting model isochrones to six O-type stars in the HR diagram. We find good fits for isochrones of 1 to 2.7\,Myr and distances of 1.7$\pm$0.2\,kpc. This puts the NGC\,6357 complex at the same distance as the NGC\,6334 complex, its close neighbor on the sky, suggesting that both clouds are physically related. Other evidence in the literature supports this notion. 
 
Using high resolution HST imaging we find that the massive star Pis\,24-18 is a binary system. Positional information suggests that the secondary component is responsible for the X-ray emission detected from this system. 
 
We have performed an optical imaging and spectroscopic survey of  Pismis\,24, the main cluster in the NGC\,6357 region, using X-ray observations to identify the cluster members. Using a combination of optical photometry and spectroscopy we estimate stellar masses, ages, and foreground extinction values of the {\newrev brightest} part of the population (the ``spectroscopic sample''). Using the well defined extinction distribution observed in the spectroscopic sample we have derived the mass and age probability function of a much larger sample of young stars for which we have good photometry but that were too faint for spectroscopy. We find that the cluster mass function of Pismis\,24 closely resembles the IMF of the Trapezium cluster down to the completeness limit of our data of $\sim$0.4\,\Msun. The median age of the cluster members is approximately 1\,Myr. 
 
We have detected five proplyds in the Pismis\,24 cluster, four of which were previously unknown. The massive stellar system Pis\,24-1 is the main source responsible for four of the proplyds. The fifth proplyd shows two tails: one pointing away from Pis\,24-1 and the other from Pis\,24-2, suggesting that this object is being photoevaporated from two directions simultaneously. Adopting a distance of 1.7\,kpc for Pismis\,24 cluster we estimate the projected distances of these proplyds from their ionizing sources to be $\sim$0.4-0.5\,pc. 
 
We employed $H-K_{\rm s}$ vs. $H-$[3.6] and $H-K_{\rm s}$ vs. $H-$[4.5] color-color diagrams to statistically investigate the disk frequency in the Pismis\,24 cluster. We find that the disk frequency in Pismis\,24 is much lower than that in clusters of similar age but without extremely massive stars. Three other clusters harboring extremely massive stars also show comparatively low disk frequencies. The dissipation timescale for the inner disks regions in massive clusters like Pismis\,24 is only roughly half of the value observed in the clusters/star-forming regions without extremely massive stars. {\newrev We discussed possible scenarios to explain the low disk frequency in Pismis\,24. We concluded that stellar encounters cannot be the main mechanism responsible for destroying disks in Pismis\,24, and argued that a combination of photoevaporation {\newnewrev and} irradiation with ionizing UV photons from nearby massive stars, causing increased MRI-induced turbulence and associated accretion activity, play an important role in the dissipation of low-mass star disks in Pismis\,24.} 
 
We also find that the disk frequency depends on the location of objects within the Pismis\,24 cluster: within 0.6\,pc from the dominant massive stellar system Pis\,24-1 the disk frequency is substantially lower than at larger radii. This provides direct observational evidence that extremely massive stars can affect the evolution of disks around intermediate- and low-mass stars in their vicinity. The observed disk frequency increases with decreasing stellar mass.

The disk frequency as traced by excess emission in the IRAC [3.6] and [4.5] bands only decreases more quickly with cluster age than the disk frequency derived when including also the [5.8] and [8.0] bands. The difference is due to the transition disk population that shows no infrared excess in the short-wavelength IRAC bands but does have excess emission in the long-wavelength bands. From these observations we derive the occurrence of transition disks among the total disk population, as a function of time, which agrees well with previously published statistics. 
 
\begin{acknowledgements} 
 Many thanks to M. Gennaro for providing the model colors of stars for 2MASS and Spitzer bands, and to the anonymous referee for comments that help to improve this paper. MF acknowledges the support by NSFC through grants 10733030 and 11173060. ASA acknowledges the support by the Deutsche Forschungsgemeinschaft (DFG), grant SI 1486/1-1. RRK acknowledges the support of an STFC studentship and post-doctoral support from a Leverhulme research project grant (F/00 144/BJ). This research has made use of the SIMBAD database, operated at CDS, Strasbourg, France. This publication makes use of data products from the Two Micron All Sky Survey, which is a joint project of the University of Massachusetts and the Infrared Processing and Analysis Center/California Institute of Technology, funded by the National Aeronautics and Space Administration and the National Science Foundation. This work is in part  based  on observations made with the Spitzer Space Telescope, which is operated by the Jet Propulsion Laboratory, California Institute of Technology under a contract with NASA. This research is  based on observations made with the NASA/ESA Hubble Space Telescope, and obtained from the Hubble Legacy Archive, which is a collaboration between the Space Telescope Science Institute (STScI/NASA), the Space Telescope European Coordinating Facility (ST-ECF/ESA) and the Canadian Astronomy Data Centre (CADC/NRC/CSA). 
\end{acknowledgements}

 \begin{appendix} 
{\newrev 
\section{Estimating  masses and ages of Pismis\,24 cluster members without spectra}\label{Appen:model}

\begin{figure*}[ht] 
\centering 
\includegraphics[width=1.\columnwidth]{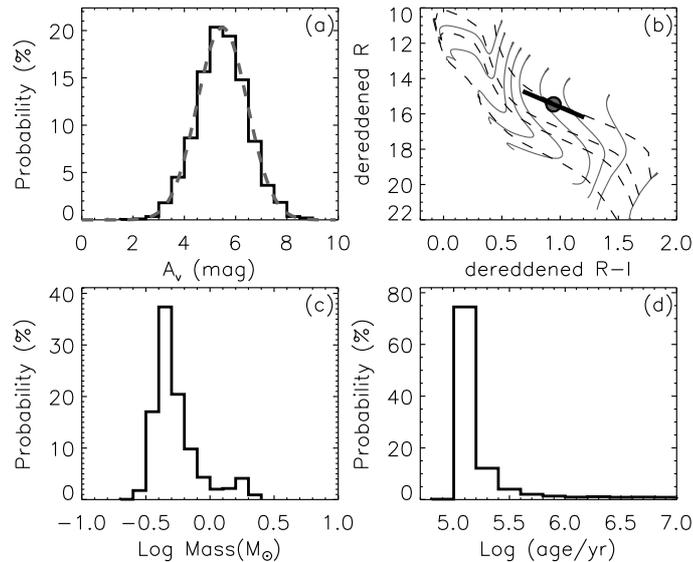} 
\caption{(a) The distribution of extinction sampled from the extinction probability function derived from \fig\,\ref{Fig:dis_spt}(b). The dashed line shows the  
the extinction probability function.  (b) The locations of source ID\#14(see \tab~\ref{tab:photometry}) on $R$~vs.~$R-I$ color-magnitude diagram. The source ID\#14 has been dereddened with the extinction values in panel (a). The filled circle show the location of source ID\#14 with highest probability. The black thick line represent the positions where source ID\#14 is located with $>$68.3\% probability. The dashed lines are PMS isochrones of  0.1, 1, 3, and 30\,Myr \citep{2008ApJS..178...89D}. The thin gray lines are the evolutionary tracks for PMS stars with  masses of 0.1, 0.3, 0.4, 0.5, 0.7, 0.9, 1.2, 2.0, and 3.0\,\Msun, respectively. The mass and age distributions of source ID\#14 derived from its locations on $R$~vs.~$R-I$ color-magnitude diagram are shown in panel (c) and (d), respectively.}\label{Fig:model} 
\end{figure*}

We cannot accurately estimate masses and ages for cluster members without proper spectral typing because we do not know the extinction towards these individual objects. In this Appendix, we explain how we derive the mass and age probability functions for those stars. We use source ID\#14 as an example. To estimate its mass and age, we we first obtained 2000 random samples from the extinction probability function derived from \fig\,\ref{Fig:dis_spt}(b). In Fig~\ref{Fig:model}(a) we show the distribution of sampled extinctions. Then, we de-reddened the photometry of source ID\#14 with each of the 2000 extinctions, and obtained 2000 positions on the $R$~vs.~$R-I$ color-magnitude diagram. In Fig~\ref{Fig:model}(b) we show the range of locations within which source ID\#14 lies with 68.3\% probability. From each position of source ID\#14 on the $R$~vs.~$R-I$ color-magnitude diagram, we estimate the mass and age through comparison to the PMS evolutionary tracks from \citet{2008ApJS..178...89D}. Repeating the $R$~vs.~$R-I$ CMD placement 2000 times we obtain 2000 mass and age estimates that are consistent with the observed photometry of source ID\#14 and with the statistical extinction distribution of the cluster, and we obtain the mass and age probability function for this object. In {\newnewrev \fig\,\ref{Fig:model}(c) and (d)}, we show the resulting mass and age probability functions. The age probability function of source ID\#14 shows a much narrower distribution than its mass probability function because the reddening vector in the $R$~vs.~$R-I$ color-magnitude diagram is approximately parallel to the PMS isochrones.}

\section{Disk frequencies in different star formation regions}\label{Appen:SFR} 
In table~\ref{tab:disk_frequency}, we list each SFR used in Fig~\ref{Fig:disk_frequency}. In the table, there are 18 SFRs, of which median ages range from $\sim$1-11\,Myr. To allow a comparative study on disk frequencies of different SFRs, we only select YSOs with masses $\ge$0.5\,\Msun \ in each SFR since the disk frequencies show dependence on stellar masses. To do this in each SFR, we use $H$-band magnitudes as a measure of YSO masses. For each SFR, we estimate the $H$-band magnitude for a 0.5\,\Msun\ PMS star with the median age of the SFR from the PMS evolutionary tracks of \citet{2008ApJS..178...89D}, and use this $H$-band magnitude as lower limit to select YSOs in that SFR. The $H$-band limit magnitude for each SFR is listed in Table~\ref{tab:disk_frequency}. We obtain photometry of each YSO in $H$, $K_{\rm s}$, [3.6], and [4.5] bands. The $H$ and $K_{\rm s}$ band photometry are from the 2MASS survey, and the photometry in IRAC [3.6] and [4.5] bands from the literature and our work. The references for that the IRAC [3.6] and [4.5] photometry come from are listed in Table~\ref{tab:disk_frequency}. For the four SFRs, ONC, OMC\,2/3, Mon\,R2, and NGC\,2244, their photometry in IRAC bands come from our work. For these SFRs, we download BCD images from Spitzer archive, mosaiced the images with Mopex software, and performed psf-fitting photometry on the mosaiced images with IDL codes described in \citet{2009A&A...504..461F}. Similar to what we have done for Pismis\,24 cluster, we estimate the disk frequency of each SFR using $H-K_{\rm s}$ vs. $H-$[3.6], and  $H-K_{\rm s}$ vs. $H-$[4.5] color-color diagrams. The YSO is considered as having an inner disk if it shows excess emission in [3.6] and [4.5] bands. The resulting disk frequency of each SFR is listed in table~\ref{tab:disk_frequency}. 

\begin{table*}
\caption{\label{tab:disk_frequency} {\newnewrev The} fractions of YSO with  [3.6] and [4.5] band excess {\newnewrev emissions} in different SFRs. Column 3: the median age of each SFR. Column 4: the references for the median age of each SFR. Column 5: the spectral types for the most massive star in each SFR. Column 6: the criteria for YSO selection. X: stars with X-ray emission, IR: stars with excess emission at infrared bands, H$\alpha$: H$\alpha$ photometry, Sp: members confirmed by spectroscopy, PM: photometric members. Colomn 7: the references for YSO selections. 
Column 8: the fractions of stars with excess emissions in [3.6] and [4.5] bands. Column 9:  $H$-band magnitude limit {\newrev for a PMS star with a mass of 0.5\,\Msun, and the corresponding age in each SFR, assuming a visual extinction of 3\,mag}. Column 10: the references for Spitzer IRAC data.
}
 \centering
\begin{tabular}{lcccccccccc}
\hline\hline
(1)              &(2)     &(3)    &(4)        &(5)          &(6)                     &(7)  &(8)        &(9)       &(10)        \\
                 & Distance    &Age    &           &             &YSO                     &     &DF         &   Limit  &            \\
Name             & (pc)   &(Myr)  &Ref        &Spt          &criteria                &Ref  &(\%)       &   (mag)  & IRAC       \\
\hline                                                                           
Taurus           &140     &  1.5  &(1)        &...          &   X, IR                &(2)  &54$\pm$8   &  9.4     &(2)         \\
IC\,348          &315     &  2.5  &(3)        &B5\,V        &   X, SP                &(3)  &32$\pm$6   &  11.5    &(3)         \\
NGC\,2068/71     &450     &  0.9  &(4)        &B1.5\,V      &   SP                   &(5)  &78$\pm$16  &  11.3    &(4, 5)    \\
ONC              &450     &  1.0  &(6)        &O6\,V        &   X                    &(7)  &63$\pm$4   &  11.6    &(8)         \\ 
OMC\,2/3         &450     &  3.0  &(9)        &...          &   X                    &(10) &39$\pm$6   &  11.9    &(8)         \\ 
Ori\,1b          &440     &  5.0  &(11, 12)   &O9.5\,V      &   SP                   &(13) &6$\pm$2    &  12.8    &(13)        \\
25\,Ori          &330     &  8.0  &(11, 12)   &B1\,V        &   SP                   &(13) &3$\pm$2    &  12.6    &(13)        \\
$\lambda$\,Ori   &400     &  5.0  &(14)       &O8\,III      &   OP, IR, SP           &(14) &4$\pm$3    &  12.6    &(14)        \\
NGC\,7160        &900     &  11.0 &(15)       &B1 II-III    &   SP                   &(15) &4$\pm$2    &  15.0    &(16)        \\
Cep\,OB\,3b      &725     &  2.5  &(17)       &O7 V         &   X                    &(17) &33$\pm$5   &  13.0    &(17)        \\
Cep\,B+S155      &725     &  1.5  &(17)       &O7 V         &   X                    &(17) &42$\pm$11  &  13.4    &(17)        \\
NGC\,2264        &800     &  3.1  &(18)       &O7 V         &   X, H$\alpha$, IR, Sp &(19) &37$\pm$3   &  13.6    &(19)        \\
NGC\,2362        &1480    &  5.0  &(20)       &O9 Ib        &   X, IR, SP            &(21) &6$\pm$2    &  15.5    &(21)        \\
Mon\,R2          &830     &  2.0  &(22)       &...          &   X                    &(23) &39$\pm$6   &  13.4    &(8)         \\
NGC\,2244        &1400    &  2.0  &(24)       &O5 V         &   X                    &(25) &22$\pm$4   &  14.5    &(8)         \\ 
NGC\,6611        &2000    &  1.0  &(26)       &O5 V         &   X                    &(26) &33$\pm$3   &  14.8    &(27)        \\ 
Cha\,I           &165     &  2.6  &(28)       &             &   SP, IR               &(29) &50$\pm$13  &  10.0     &(29)        \\ 
$\sigma$\,Ori    &440     &  3.0  &(30)       &O9.5 V       &  X, IR, SP, PM         &(30, 31) &25$\pm$4  &  12.4    &(30)        \\ 
NGC\,7129        &1000    &  3.0  &(31)       &...          &  X                     &(32) &29$\pm$9  &  14.2    &(32)        \\ 
$\gamma$\,vel    &350     &  5.0  &(33)       &WC           &  X, PM                 &(33) &0.5$\pm$0.4&  12.3    &(33)        \\ 
Tr\,37           &900     &  4.0  &(15)       &O6.5\,V      &  SP, X                 &(15,34)&42$\pm$5 &  14.2    &(16,34)        \\
\hline
\normalsize
\end{tabular}
\tablebib{(1) \citet{2002ApJ...580..317B}; (2) \citet{2010ApJS..186..111L}; (3) \citet{2006AJ....131.1574L}; 
 (4) \citet{2009A&A...504..461F}; (5) \citet{2008AJ....135..966F}; (6) \citet{1997AJ....113.1733H}; (7) \citet{2005ApJS..160..319G}; (8) this paper; 
(9) \citet{2005PhDT.........8P}; (10) \citet{2002ApJ...566..974T}; (11) \citet{2005AJ....129..907B}; (12) \citet{2007ApJ...661.1119B};
 (13) \citet{2007ApJ...671.1784H}; (14) \citet{2007ApJ...664..481B}; (15) \citet{2005AJ....130..188S}; (16)\citet{2006ApJ...638..897S};
(17) \citet{2009ApJ...699.1454G}; (18) \citet{2004AJ....128.1684S}; 
(19) \citet{2009AJ....138.1116S}; (20) \citet{2001ApJ...563L..73M}; (21) \citet{2009ApJ...698....1C}; 
(22) \citet{1997AJ....114..198C}; (23) \citet{2002ApJ...567..423K}； (24) \citet{2002AJ....123..892P}; (25) \citet{2008ApJ...675..464W}; 
(26) \citet{2007A&A...462..245G}; (27) GLIMPSE archive data; (28) \citet{2004ApJ...602..816L,2007ApJS..173..104L}; (29) \citet{2008ApJ...675.1375L}; 
(30) \citet{2008ApJ...688..362L}; (31) \citet{2007ApJ...662.1067H}; (32) \citet{2009A&A...507..227S}; (33) \citet{2008ApJ...686.1195H}. (34) \citet{2009AJ....138....7M}}
\end{table*}

\end{appendix} 
 
\bibliographystyle{aa} 
\bibliography{references}

\renewcommand{\tabcolsep}{0.05cm}
\tiny
\onltab{1}{\longtab{1}{
\begin{landscape}
\begin{longtable}{lccccccccccccccc}
\caption{\label{tab:photometry}Photometric magnitudes for stars with X-ray emission in PISMIS\,24.  Column 2: the identification numbers in \citet{2007ApJS..168..100W}. Column 3: the distances between xray sources and optical/infrared counterparts. }\\
\hline\hline
(1)&(2)& (3) & (4) & (5) & (6) &  (7) & (8) &(9) & (10) &(11) &(12) &(13)&(14)\\
& &dis & RA & DEC & $R$ & $I$ &  $J$ & $H$ &$K_{s}$ & [3.6] &[4.5] &[5.8] &[8.0]\\
ID& W07&(arcsec) & (J2000) & (J2000) & (mag) & (mag) & (mag) & (mag) &(mag)& (mag) & (mag) &(mag)  & (mag)\\
\hline
\endfirsthead
\caption{continued.}\\
\hline\hline
(1)&(2)& (3) & (4) & (5) & (6) &  (7) & (8) &(9) & (10) &(11) &(12) &(13)&(14)\\
& & & RA & DEC & $R$ & $I$ &  $J$ & $H$ &$K_{s}$ & [3.6] &[4.5] &[5.8] &[8.0]\\
ID& W07&(arcsec)&(J2000) & (J2000) & (mag) & (mag) & (mag) & (mag) &(mag)& (mag) & (mag) &(mag) & (mag) \\
\hline
\endhead
\hline
\endfoot
1&1&0.81& 17 23 53.93& -34 08 50.5& 22.793$\pm$0.080& 21.572$\pm$0.029&...&...&...&...&...&...&...&\\
2&2&0.48& 17 23 58.39& -34 08 02.4& 21.952$\pm$0.082& 19.076$\pm$0.022& 15.18$\pm$0.05& 13.18$\pm$0.04& 11.93$\pm$0.03& 10.69$\pm$0.04& 10.49$\pm$0.06& 10.13$\pm$0.13&...&\\
3&3&0.87& 17 23 59.16& -34 12 16.9& 15.595$\pm$0.001& 14.080$\pm$0.001& 11.93$\pm$0.02& 11.18$\pm$0.03& 10.69$\pm$0.03&...&...&...&...&\\
4&6&0.32& 17 24 03.22& -34 14 02.2& 16.767$\pm$0.001& 14.949$\pm$0.001& 12.67$\pm$0.02& 11.68$\pm$0.02& 11.31$\pm$0.02& 11.01$\pm$0.05& 10.96$\pm$0.07& 11.23$\pm$0.23&...&\\
5&8&0.45& 17 24 05.46& -34 12 57.5&...&...&...&...&...& 12.61$\pm$0.08& 11.70$\pm$0.08& 11.23$\pm$0.18&...&\\
6&9&0.53& 17 24 05.61& -34 07 09.7&...&...& 9.03$\pm$0.02& 8.63$\pm$0.05& 8.46$\pm$0.02& 8.35$\pm$0.06& 8.33$\pm$0.12& 8.27$\pm$0.14&...&\\
7&10&0.50& 17 24 06.66& -34 13 05.1&...&...&...& 14.52$\pm$0.07& 13.32$\pm$0.05& 12.66$\pm$0.12& 12.31$\pm$0.21&...&...&\\
8&11&0.65& 17 24 06.81& -34 11 49.2&...& 22.390$\pm$0.128& 16.53$\pm$0.12& 13.83$\pm$0.03& 12.38$\pm$0.03& 11.05$\pm$0.04& 10.61$\pm$0.05& 10.22$\pm$0.06& 9.92$\pm$0.09&\\
\end{longtable}
\end{landscape}
} }
\normalsize

\renewcommand{\tabcolsep}{0.05cm}
\tiny
\onltab{3}{\longtab{3}{
\begin{longtable}{lccccccccccccccc}
\caption{\label{tab:mass_age}Stellar parameters for X-ray emissing stars with spectral types in PISMIS\,24.  Column 2: the identification numbers in \citet{2007ApJS..168..100W}}\\
\hline\hline
(1)&(2)& (3) & (4) & (5) & (6) &(7) &(8)\\
&  & RA & DEC &  & $A_{\rm v}$ &mass &age\\
ID& W07 & (J2000) & (J2000) & Spt & (mag) & ($M_{\odot}$) &(Myr) \\
\hline
\endfirsthead
\caption{continued.}\\
\hline\hline
(1)&(2)& (3) & (4) & (5) & (6) &(7) &(8)\\
&  & RA & DEC &  & $A_{\rm v}$ &mass &age\\
ID& W07 & (J2000) & (J2000) & Spt & (mag) & ($M_{\odot}$) &(Myr) \\
\hline
\endhead
\hline
\endfoot
3&       3& 17 23 59.16& -34 12 16.9&G5&4.369&2.938&1.4\\
4&       6& 17 24 03.22& -34 14 02.2&G5&5.508&2.665&1.9\\
9&      12& 17 24 07.66& -34 17 52.4&K5&6.440&0.862&0.6\\
15&      18& 17 24 13.60& -34 09 21.3&K2&3.661&1.540&1.1\\
16&      19& 17 24 13.60& -34 16 56.7&M9&7.492&...&...\\
18&      21& 17 24 13.74& -34 06 57.1&M9&19.531&...&...\\
19&      22& 17 24 14.72& -34 10 43.7&F8&1.798&1.664&8.1\\
25&      30& 17 24 16.18& -34 13 11.4&K2&2.634&1.475&2.9\\
\end{longtable}
} }
\normalsize

\end{document}